\documentclass{aa}  
\usepackage{natbib}

\usepackage{graphicx}
\usepackage{txfonts}
\usepackage{multirow}
\usepackage{booktabs}
\usepackage{hyperref}

\newcommand{\LEt}[1]{{\bf}}

\makeatletter
\renewcommand*\aa@pageof{, page \thepage{} of \pageref*{LastPage}}
\begin{document}

   \title{Testing ultralow amplitude Cepheid candidates in the Galactic disk by TESS and Gaia}
      \titlerunning{Testing ultralow amplitude Cepheid candidates in the Galactic disk}


   \author{Dóra Tarczay-Nehéz
          \inst{1,2,3}\fnmsep\thanks{e-mail: tarczaynehez.dora@csfk.org}
          \and
          László Molnár\inst{1,2,3,4}
          \and 
          Attila Bódi\inst{1,2,3}
          \and Róbert Szabó\inst{1,2,3,4}     
          }

   \institute{CSFK, MTA Centre of Excellence, H-1121 Budapest, Hungary
             \and
             Konkoly Observatory, Research Centre for Astronomy and Earth Sciences, E\"otv\"os Lor\'and Research Network (ELKH), Konkoly Thege Mikl\'os \'ut 15-17, H-1121 Budapest, Hungary
              \and
              MTA CSFK Lend\"ulet Near-Field Cosmology Group
             \and 
              E\"otv\"os Lor\'and University, Institute of Physics, H-1117 P\'azm\'any P\'eter s\'et\'any 1/A, Budapest, Hungary
             }
 \authorrunning{D. Tarczay-Nehéz et al.}
   \date{Received XX; accepted XX}

 
  \abstract{Ultralow amplitude (ULA) and strange mode Cepheids are thought to be pulsating variable stars that are near to or are at the edges of the classical instability strip. Until now, a few dozen such variable star candidates have been found both in the Large Magellanic Cloud and the Milky Way. For the present work, we studied six ULA Cepheid candidates in the Milky Way, identified by using CoRoT and 2MASS data. In order to identify their positions in the period--luminosity and color--magnitude diagrams, we used the Gaia DR3 parallax and brightness data of each star to calculate their reddening-free absolute magnitudes and distances. Furthermore, we calculated the Fourier parameters (e.g., period and amplitude) of the light variations based on CoRoT and TESS measurements, and established the long-term phase shifts for four out of six stars. 
  Based on the results, we conclude that none of the six ULA Cepheid candidates are pulsating variable stars, but rather rotation-induced variable stars (rotational spotted and $\alpha^2$~Canum Venaticorum variables) that are either bluer or fainter than Cepheids would be.
  }

  
   


   \maketitle
%

\section{Introduction}

   Many classical pulsating stars are characterized by significant periodic or multi-periodic brightness variations that exceed several hundredths to a few tenths of relative magnitudes. In contrast, ultralow amplitude (ULA) pulsations are usually referred to as millimagnitude-level changes in stars that would otherwise show normal pulsations as well. A subgroup of ULA pulsating modes are strange modes that were first proposed to exist in highly nonadiabatic luminous stars, for example, post-asymptotic giant branch (post-AGB) stars, by \cite{Woodetal1976}. 
   The term "strange mode" was first proposed by \cite{Coxetal1980}, investigating the pulsation modes of R~Coronae Borealis stars.
   \cite{Buchleretal1997} showed that strange modes are surface modes and can also exist in, for example, classical Cepheids, even though they are weakly nonadiabatic stars. Thus, strange modes should occur at the adiabatic limit as well. 
   
   \cite{Buchleretal1997} and \cite{BuchlerandKollath2001} found that strange modes can be linearly unstable leftward of the hot blue edge of the instability strip.  
   The latter work also pointed out that strange modes appear when one-quarter of a wavelength fits approximately into the surface layer of the pulsating envelope above the hydrogen partial ionization zone. 
   In this case, the mode is trapped in the outermost region where it can decouple from the inner regions and the pulsation mode becomes dominant over the other modes.
   Hydrodynamic calculations revealed that the surface radial velocities of strange modes (being ULA modes) fall within the  0.1 – 1.0 km/s range, and the luminosity variations are expected to be on the order of millimagnitudes. Thus, strange modes are difficult to observe.

\begin{table*}
\caption{Basic CoRoT data of the ULA candidates.}
\label{table:corot_data}      
\centering            
\begin{tabular}{c c c c c c c c  }
\hline\hline            
CoRoT & $\alpha$ & $\delta$& $W$ & $\sigma_W$ & Frequency & Period & Amplitude \\ 
ID & (J2000) & (J2000) & (mag) & (mmag) & [c/d] & (days) & (mmag)  \\
\hline
10283{\bf5671} & 101.74965 &-0.51000 & 15.347 & 7.535 & $f_0 = 0.8372$ & 1.1945 & 11.94 \\
& &&&& $2f_0$ & & 0.98 \\
& &&&& $3f_0$ & & 0.28 \\
10296{\bf4858}\tablefootmark{a} &102.46044 & -2.46076 & 14.087 & 7.378 & $f_0 = 0.3911$ & 2.5572 & 10.67 \\
&&& & &$3f_0$ & & 1.14 \\
10294{\bf0813}\tablefootmark{a} & 102.33850 & -1.82978 & 13.905 & 9.378 & $f_0 = 0.2681$ & 3.7293 & 16.47 \\
&&&&& $2f_0$ & & 3.72 \\
10283{\bf8469}\tablefootmark{a} & 101.76825 & -1.20753 & 12.549 & 12.767 & $f_0 = 0.1771$ & 5.6481 & 12.14 \\
&&&&& $2f_0$ && 1.55 \\
10282{\bf7671} & 101.69550 & -2.52647 & 14.380 & 4.661 & $f_0 = 0.1163$ & 8.6021 & 12.36 \\
&&&&& $2f_0$ && 3.46 \\
&&&&& $ 3 f_0$ && 0.61 \\
10279{\bf9143} & 101.48186 & -2.29820 & 15.938 & 11.412 & $f_0 = 0.0674$ & 14.8434 & 28.36 \\
&&&&& $2f_0$ && 5.49 \\

\hline                                  
\end{tabular}
\tablefoot{White magnitude ($W$) and its error ($\sigma_W$) data from the CoRoT AsteroSeismology (AS) channel \citep{Auvergneetal2009}. 
\tablefoottext{a}{Frequency and amplitude values indicated in the table were calculated based on red channel data by \citet{Szaboetal2009}.}
}
\end{table*}

\label{sec:corot_data}
\begin{table*}
\caption{2MASS data about the six ULA candidates.}  
\label{table:2mass_data}      
\centering            
\begin{tabular}{c c c c c c c c}
\hline\hline            
\multirow{3}{*}{CoRoT ID} &  \multicolumn{7}{c}{2MASS}\\ 
\cmidrule(lr){2-8}
&  \multirow{2}{*}{ID}& $J$ &$\sigma_J$ & $H$ & $\sigma_H$ & $K$ & $\sigma_K$\\
&  &  (mag) & (mag) & (mag)&  (mag) & (mag) & (mag)\\
\hline
   102835671 & 06465991-0030360 & 14.136 & 0.026 & 13.894 & 0.041 & 13.766 & 0.052\\
   102964858 & 06495050-0227387 & 13.273 & 0.023 & 13.086 & 0.022 & 13.011 & 0.026\\
   102940813 & 06492123-0149471 & 13.126 & 0.023 & 13.044 & 0.029 & 12.946 & 0.037\\
   102838469 & 06470437-0112269 & 11.903 & 0.023 & 11.838 & 0.023 & 11.800 & 0.024\\
   102827671 & 06464691-0231353 & 13.095 & 0.024 & 12.614 & 0.032 & 12.515 & 0.029\\
   102799143 & 06455564-0217534 & 14.672 & 0.03 & 14.372 & 0.038 & 14.385 & 0.087\\

\hline                                  
\end{tabular}
\end{table*}

   In terms of pulsation, the amplitude growth time, $\tau_{\mathrm{gr}}$, is much smaller than the stellar evolutionary time, $\tau_{\mathrm{evol}}$. 
   Thus, inside the instability strip, the amplitude growth rate, $\eta \sim \tau_{\mathrm{gr}}^{-1}$, which describes the change of the fractional change in the mode's kinetic energy per period, is positive. 
   Inside the strip, pulsation amplitudes grow and then saturate very quickly compared to stellar evolution.
   However, at the edges of the instability strip, $\eta$ equals zero, and thus the relation becomes $\tau_{\mathrm{gr}} \gg \tau_{\mathrm{evol}}$. 
   Hence, at the vicinity of the instability strip, pulsation does not occur immediately at a full pulsational amplitude, but it stays at a low amplitude for a time on the order of $(\tau_{\mathrm{gr}}\tau_{\mathrm{evol}})^{1/2}$ and then grows up to full amplitude quickly \citep{Buchleretal2005}.
   Therefore, in terms of observations, one would expect a gap in the number of Cepheids between ULA Cepheids and classical Cepheids since, at the stage of rapid amplitude growth, very few Cepheids are observable. 
   
   Despite the difficulties as to the observability, a few ULA and strange mode Cepheid candidates were already found both in the Large Magellanic Cloud and in the Milky Way. 
   The first candidates in the Large Magellanic Cloud were published by \cite{Buchleretal2005} based on Massive Compact Halo Object (MACHO) and Optical Gravitational Lensing Experiment (OGLE-II, and OGLE-III) data. 
   They found 14 Cepheids with amplitudes in the MACHO $M_\mathrm{R}$ band below 50 mmag, seven of which have amplitudes of only $6$ mmag. 
   \cite{Buchleretal2005} refer to the latter seven stars as ULA Cepheids.
   They also found two additional low-amplitude Cepheids that may have strange modes, and thus they refer to those as strange mode Cepheid candidates. 
   Additional ULA Cepheids have been reported in the Magellanic Clouds by, for example, \cite{Soszynskietal2008} and \cite{Buchleretal2009}.

   In the Milky Way, \cite{Szaboetal2009} identified 37 ULA Cepheid candidates using data from the Convection, Rotation, and planetary Transits (CoRoT) space telescope \cite{corot-2006}. 
   In that paper, light curves and Fourier parameters were presented for the six best candidates from the initial run in the Galactic anticenter direction (IRa01) dataset.
   They showed that these stars lay in the vicinity of the blue and red edges of the Cepheid instability strip on the period\,--\,($J-H$) color index plane (in the "strange" region).
   As the data of the remaining 31 are unpublished (including their CoRoT IDs), we were only able to investigate those six stars in this work, by means of Transiting Exoplanet Survey Satellite (TESS) and Gaia data \citep{Gaiamission2016}.


\section{Data selection and processing}
\label{sec:data}

\subsection{CoRoT and 2MASS data}
\label{sect:corot}

\begin{table*}
\caption{Gaia DR3 data for ULA Cepheid candidates.}            
\label{table:uladata}      
\centering            
\resizebox{\textwidth}{!}{\begin{tabular}{c c c c c ccc cc cc c}
\hline\hline            
\multirow{3}{*}{CoRoT ID} & \multirow{3}{*}{TESS TIC} &  \multicolumn{10}{c}{Gaia}\\ 
\cmidrule(lr){3-13}
&  & \multirow{2}{*}{DR3 ID}& $G$ &$\sigma_G$ & $G_\mathrm{BP}$ & $\sigma_{G_\mathrm{BP}}$ & $G_\mathrm{RP}$ & $\sigma_{G_\mathrm{RP}}$ & $d$ &$\sigma_d$\tablefootmark{a} & $\varpi$ & $\sigma_\varpi$ \\
&  & &  (mag) &(mmag) & (mag) & (mmag)& (mag) & (mmag)& (pc) & (pc) & (mas) & (mas)\\
\hline                    
   102835671 & 36009248 & 3107394927263669120 & 15.527 & 1.319 & 15.855 &  7.766 & 14.884 &  5.993 & 6044 & $^{+1633}_{-1322}$ & 0.1179 & 0.0568\\
   102964858 & 281621267 & 3106121692798422016 & 14.185 & 0.715 & 14.425 & 2.527 & 13.784 & 2.337 & 1570 & $^{+36}_{-36}$ & 0.5930 & 0.0181\\
   102940813 & 36558785 & 3106295896673128576 & 13.944 & 0.890 & 14.167 & 3.062 & 13.566 & 2.487 & 4078 & $^{+388}_{-284}$  & 0.2033 & 0.0184\\
   102838469 & 36011203 & 3107142829863485952 & 12.622 & 0.634 & 12.803 & 2.089 & 12.291 & 1.176 & 2191 & $^{+197}_{-152}$ & 0.4322 & 0.0310\\
   102827671 & 35956145 & 3106150589338177920 & 14.493 & 0.516 & 14.984 & 2.684 & 13.835 & 1.471 & 465 & $^{+5}_{-5}$  & 2.1074 & 0.0252\\
   102799143 & 168195156 & 3106171926734960896 & 16.014 & 2.024 & 16.441 & 7.236 & 15.408 & 5.807 & 5781 & $^{+2318}_{-1354}$ & 0.1285 & 0.0516\\
\hline                                  
\end{tabular}}

\end{table*}

   In order to distinguish between low-amplitude Cepheids and other low-amplitude variables (such as spotted stars and eclipsing binaries), \cite{Szaboetal2009} used a four-step data selection process, which is described in detail in their paper, and thus it is not detailed in this work.
   Table\,\ref{table:corot_data} collects the basic data -- ID, position, white magnitude, and its error in the CoRoT AsteroSeismology (AS) channel \citep{Auvergneetal2009} -- of the ULA candidates from the CoRoT Faint Star Catalogue \citep{CorotFaint} and the calculated light variation frequencies and amplitudes by \cite{Szaboetal2009}. 
   We use the last four digits of the CoRoT IDs for identification throughout this paper.
   The final CoRoT light curve products for these stars are published in Appendix~\ref{app:photometry} in their entirety.

   We note that the frequency and amplitude values for stars 4858, 0813, and 8469 listed in Table\,\ref{table:2mass_data} were calculated from the integrated red flux data by \citet{Szaboetal2009}. 
   In these cases, we collected the chromatic red, blue, and green channel data of each star, as white flux data can be generated as a combination of chromatic channel data \citep[see more details in, e.g.,][]{CorotBook2016}. We then used the white flux values in our subsequent analysis. The $\sigma_W$ values for these stars are calculated in Table\,\ref{table:corot_data} with the pooled standard deviation method using the standard deviation values of red, green, and blue channel data (i.e., \texttt{REDFLUXDEV}, \texttt{GREENFLUXDEV}, and \texttt{BLUEFLUXDEV}).

   \citet{Szaboetal2009} only collected the $J-H$ color indices of each star from the Two Micron All Sky Survey (2MASS) database. 
   Thus, we cross-matched the positions of all six ULA candidates with the 2MASS point source catalog of \citet{Cutrietal2003} to collect the mean $J$-, $H$-, and $K$-band magnitudes of each star with the automated \texttt{python} package described in Section\,\ref{subsec:gaiadata}.
   2MASS data (ID, mean $J$, $H$, and $K$ magnitudes) are listed in Table\,\ref{table:2mass_data}.

\subsection{Photometry in TESS sectors 6 and 33}
\label{sect:tess}
   The TESS mission covered the CoRoT anticenter fields in Sectors 6 and 33 \citep{TESS2015}, including the six, yet published ULA Cepheid candidates. 
   Although some of the six candidates are faint (fainter than 15\,mag in CoRoT optical wavelengths), we were able to clearly detect the low-amplitude variations in four of them with TESS. 
   Table\,\ref{table:uladata} lists the TESS Input Catalog IDs cross-matched with CoRoT and Gaia DR3 catalogs.
   To compare light curves with CoRoT data, TESS full-frame images (FFIs) were processed utilizing the FITSH software package developed by \cite{Pál2012}. 
   Since the stars are in dense stellar environments within the disk, we used differential-image photometry to lessen the effects of blending. The steps of the data processing are described in detail by \citet{Pál2020}, as well as by \citet{Plachyetal2021} and \citet{Molnáretal2022} who used the same technique specifically for pulsating stars. This differential-image pipeline has been used for a variety of other variable stars successfully \citep[see, e.g.,][]{borkovits-2022,czavalinga-2022,SzabóZs-2021,SzabóZs-2022,zsidi2-2022,zsidi1-2022}.
   
   In short, during the data processing, we performed astrometric cross-matching with the Gaia (E)DR3 catalog \citep{gaia-edr3-2021}, and then the flux zero-point between the TESS and Gaia photometric system was derived. 
   In the subsequent step, we generated $64\times64$ pixel  image cutouts centered on the target star and then created differential images from them by subtracting a reference image from each frame.
   For the reference image, we needed frames where the deviations are expected to be minimal for all the individual images and all have a good signal-to-noise ratio. 
   We used the median average of 11 and 13 images for S06 and S33, respectively, around the mid-time of the observation series to create the reference images. 
   This way, the apparent shift of stars due to differential velocity aberration is minimal. 
   
   In order to correct many intrinsic and/or instrumental effects, image convolution was done using the reference image. 
   For this, we needed to determine the convolution coefficients with the \texttt{fiphot} task of \texttt{FITSH}. 
   We then extracted the brightness variations with aperture photometry. 
   Since the targets are faint and the field is crowded, we used a small aperture with a radius of 1.5~px.
   
   In parallel, we needed to obtain a set of reference fluxes for the targets, which is difficult to determine due to the large pixel size of TESS ($21\arcsec$px$^{-1}$). 
   Thus, we used Gaia DR3 $G_\mathrm{RP}$ (\texttt{phot\_rp\_mean\_mag}) magnitudes for each star since that passband is very similar to the passband TESS uses.
   In the final step, the fluxes were first adjusted to the mean, as the applied reference image is not necessarily averaged over the light variation of the star.
   Then we added the mean flux calculated from the Gaia $G_\mathrm{RP}$ brightness to the differential fluxes and converted them to magnitudes. The final TESS light curves are published in Appendix~\ref{app:photometry}.
   
\begin{figure*}
    \centering
    \includegraphics[width=.97\textwidth]{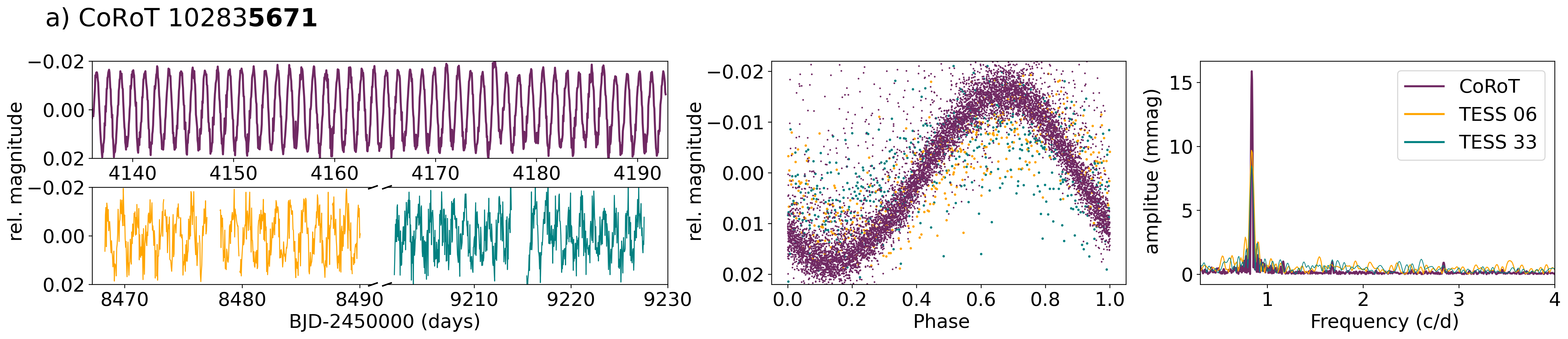}
    \includegraphics[width=.97\textwidth]{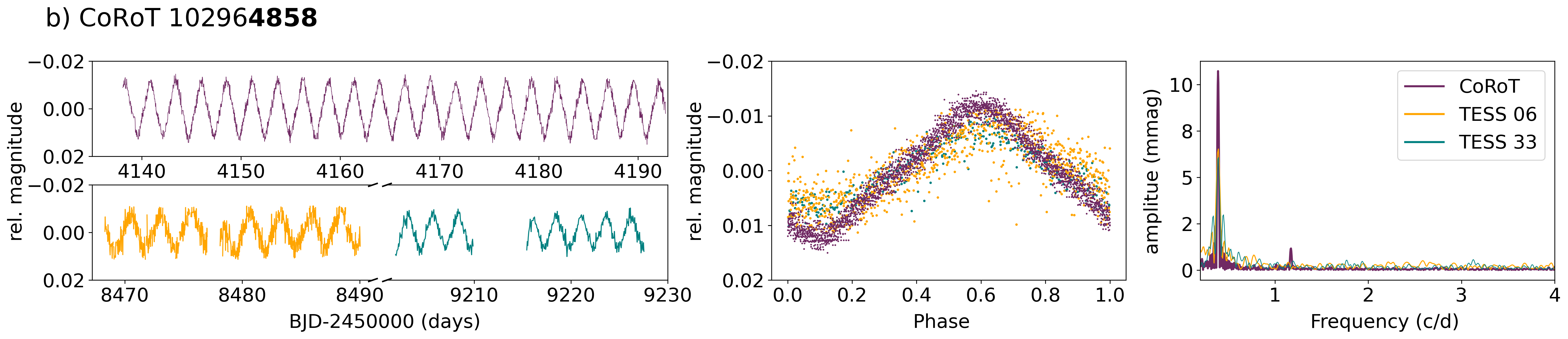}
    \includegraphics[width=.97\textwidth]{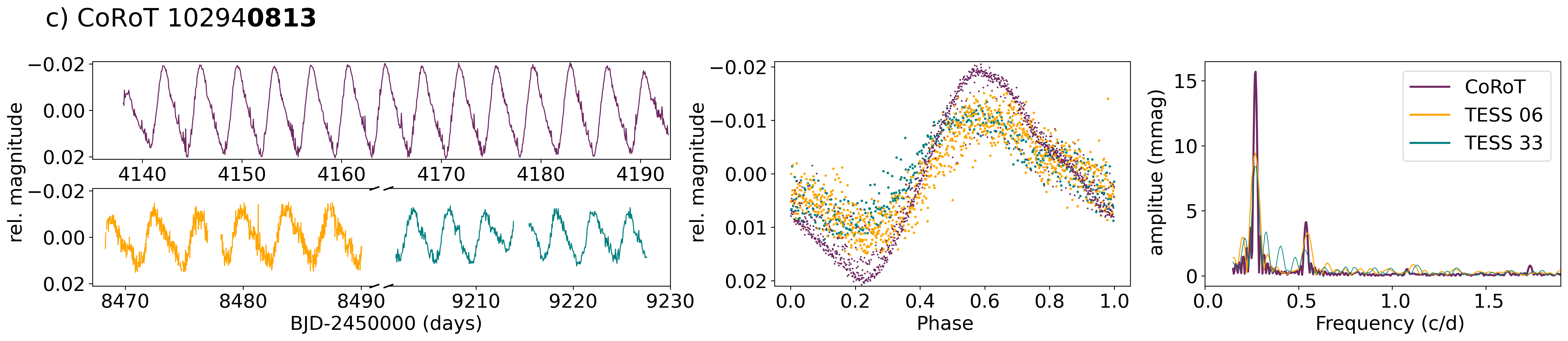}
    \includegraphics[width=.97\textwidth]{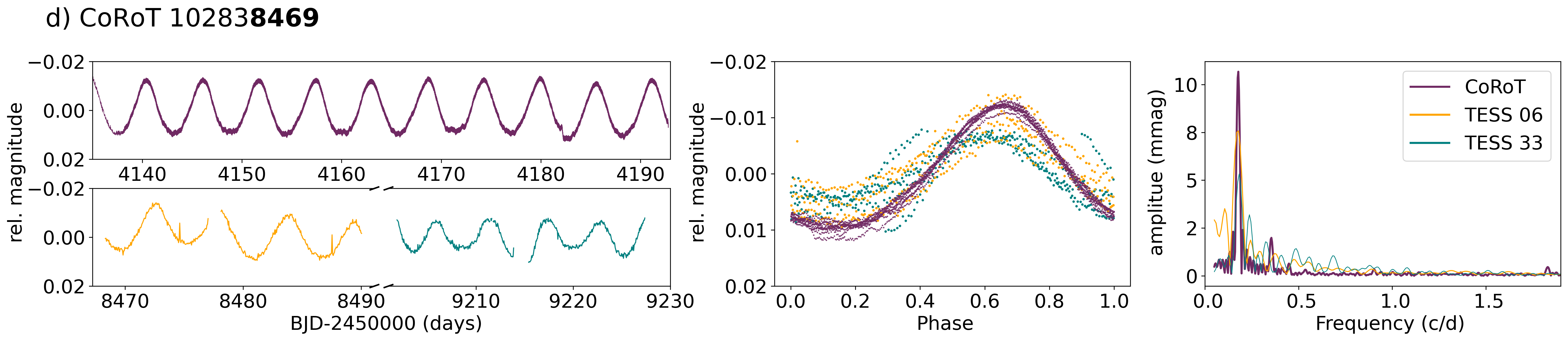}
    \includegraphics[width=.97\textwidth]{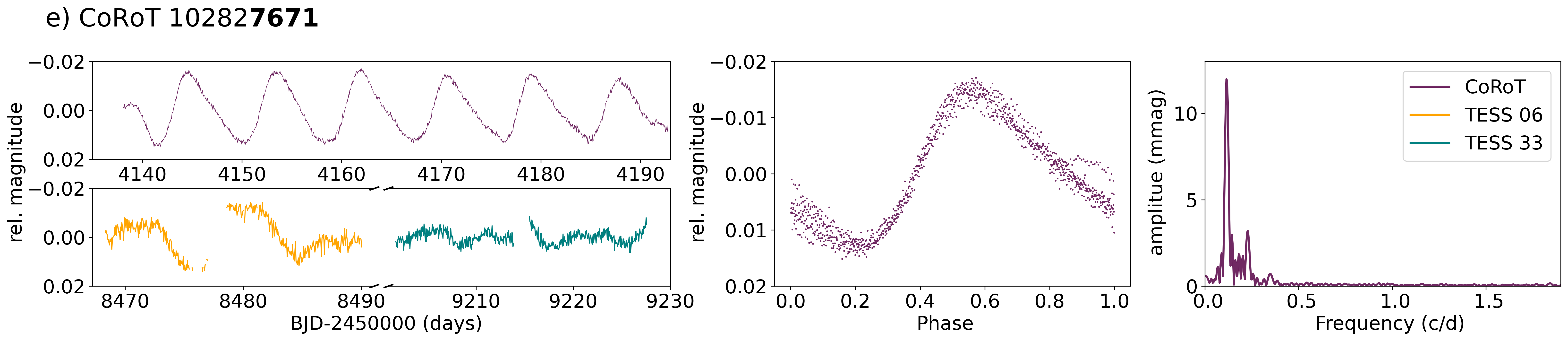}
    \includegraphics[width=.97\textwidth]{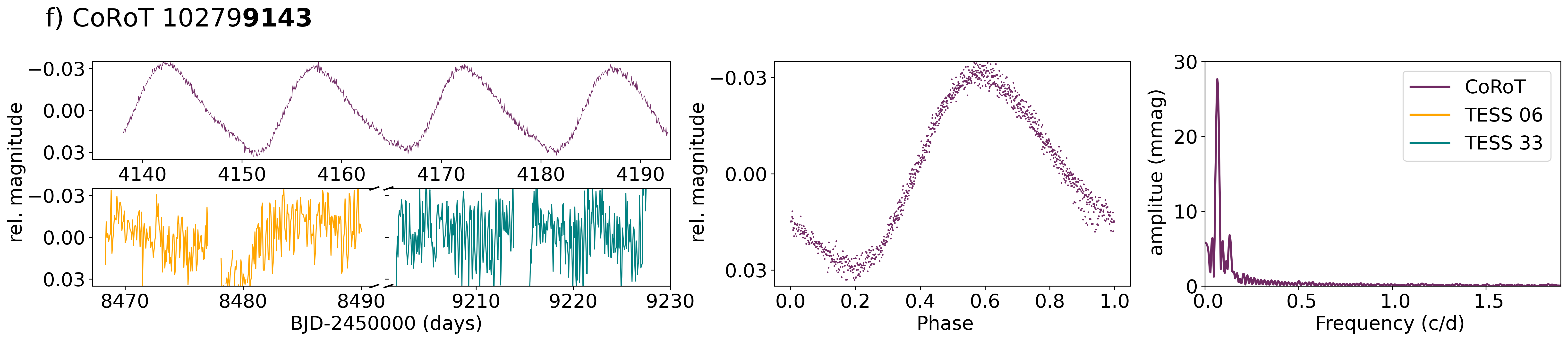}
    \caption{CoRoT (purple) and TESS (sectors 6 and 33, marked with orange and teal, respectively) measurement data (left panels) of the six stars. Folded light curves (middle panels) and power spectra (right panels) are plotted with TESS data only for those stars where we could determine their period (panels a - d).
    }
    \label{fig:all_lcs}
\end{figure*}

   The left-hand panels of Figure\,\ref{fig:all_lcs} present the CoRoT $W$ (purple) and TESS light curves from sectors 6 (orange) and 33 (teal), respectively, normalized to the mean magnitude measured in the corresponding bandpass. 
   One can easily see that the light variation of four (5671, 4858, 0813, and 8469 on panels \textit{a--d}) out of the six candidates can be found both in CoRoT and TESS data. 
   However, the light variations of stars in panels \textit{e} and \textit{f} (7671 and 9143) are either not clearly identifiable or the signal-to-noise ratio is too low due to the faintness of the star 
   and variations likely come from contaminating nearby sources.

\subsection{Fourier parameters}
\label{sec:fourierdesc}

   To calculate Fourier parameters, we first removed the outlier data points outside the $2\sigma$ levels relative to the mean.
   Then, we fit the light curves with a Fourier series in the form of 
   \begin{equation}
   m = m_0 + \sum_i A_i \sin\left( 2\pi i f_0 t +\phi_i \right),  
   \end{equation}
   \noindent where $m_0$ is the average magnitude, $f_0$ is the frequency corresponding to the highest peak in the power spectrum, while $A_i$, $\phi_i$ are the amplitude and phase of the corresponding $i^\mathrm{th}$ harmonic, respectively. 
   Then we calculated the relative Fourier parameters of each harmonic, as defined by \cite{simon-lee-1981}: 
   \begin{equation}
       R_{i1} = A_i/A_1,
   \end{equation} and 
   \begin{equation}
       \phi_{i1} = \phi_i - i\phi_1.
   \end{equation}
   Errors were calculated in the analytic form as defined by \cite{Breger1999}.
   These steps were performed both as a part of the Python-based package described in the next subsection and with the \texttt{PERIOD04} software of \cite{LenzandBreger2005}.

\begin{table*}
\caption{Calculated periods and relative Fourier parameters for the first three harmonics.}            
\label{table:res_fourier}      
\centering            
\resizebox{\textwidth}{!}{\begin{tabular}{c c c c c c c c c c c c}
\hline\hline            
CoRoT ID & $P$ & $A_1$ & $\sigma A_1$ &  $R_{21}$ & $\sigma R_{21}$ & $R_{31}$  & $\sigma R_{31}$ & $\phi_{21}$ & $\sigma\phi_{21}$ & $\phi_{31}$ & $\sigma\phi_{31}$ \\
& (days) & (mmag) & (mmag)  &  &  &  &  & (rad) & (rad) & (rad) & (rad) \\
\hline 
\noalign{\smallskip}
\multicolumn{12}{c}{{\emph{CoRoT data}}}\\\\ 
102835671 & 1.1945 & 16.6536 & 0.0607  & 0.0531 & 0.0037 & 0.0184 & 0.0036 & 3.3251 & 0.0110 & 5.1682 & 0.0317\\
102964858 &   2.5626 & 7.8442 & 0.0269 & $<$ 0.0001 & - & 0.1151 & 0.1773 & - &  -& 4.6586 & 0.0282\\
102940813 & 3.7101 & 14.3888 & 0.0576 &  0.3019 & 0.0017 & 0.0312 & 0.0016 & 5.4451 & 0.0010 & 0.2222 & 0.0109\\
102838469 & 5.6406 & 10.7955 & 0.0200 & 0.1502 & 0.0019 & 0.0302 & 0.0018 & 2.5133 & 0.0019 & 3.9384 & 0.0218\\
102827671 & 8.6404 & 11.1005 & 0.0539 &  0.3339 & 0.0051 & < 0.0001 & - & 6.0053 & 0.0028 & - &-\\
102799143 & 14.9908 &  30.3857 & 0.1018 &  0.1704 & 0.0034 & < 0.0001 &- & 5.9593 & 0.0033 & -&-\\\\

\noalign{\smallskip}
\multicolumn{12}{c}{{\emph{TESS data from sector 06}}}\\\\
102835671 & 1.1916 & 11.9808 & 7.7351 &  - & - & - & - & - & - & - & -\\
102964858 &  2.5628 & 7.3769 & 0.1780 & - & - & - & - & - & - & - & -\\
102940813 & 3.7234 & 9.7030 & 0.1226 &  0.3192 & 0.0133 & - & - & 2.9850 & 0.0133 & - & -\\
102838469 & 5.6251 & 9.0109 & 0.1030 & - & - & - & - & - & - & - &-\\
102827671 & - & - & - & - & - & - & - &  - & - & - &-\\
102799143 & - & - & - & - & - & - & - &  - & - & - &-\\\\

\noalign{\smallskip}
\multicolumn{12}{c}{{\emph{TESS data from sector 33}}}\\\\
102835671 & 1.2068 & 12.1461 & 1.3424 &  - & - & - & - & - & - & - & -\\
102964858 &   2.5641 & 7.9805 & 0.1900 & - & - & - & - & - & - & - & -\\
102940813 & 3.7188 & 9.9865 & 0.0950 &   0.2993 & 0.0099 & - & - & 2.9181 & 0.0059 & - & -\\
102838469 & 5.6611 & 9.2569 & 0.0512 & - & - & - & - & - & - & - &-\\
102827671 & - & - & - & - & - & - & - &  - & - & - &-\\
102799143 & - &  - & - & - & - & - & - &  - & - & -&-\\
\hline                                  
\end{tabular}}

\tablefoot{
 Upper limits for $R_{21}$ and $R_{31}$ values were calculated from the background noise of the Fourier power spectrum at the given harmonic's frequency.}
\end{table*}

   \begin{figure}
    \centering
    \includegraphics[width=\columnwidth]{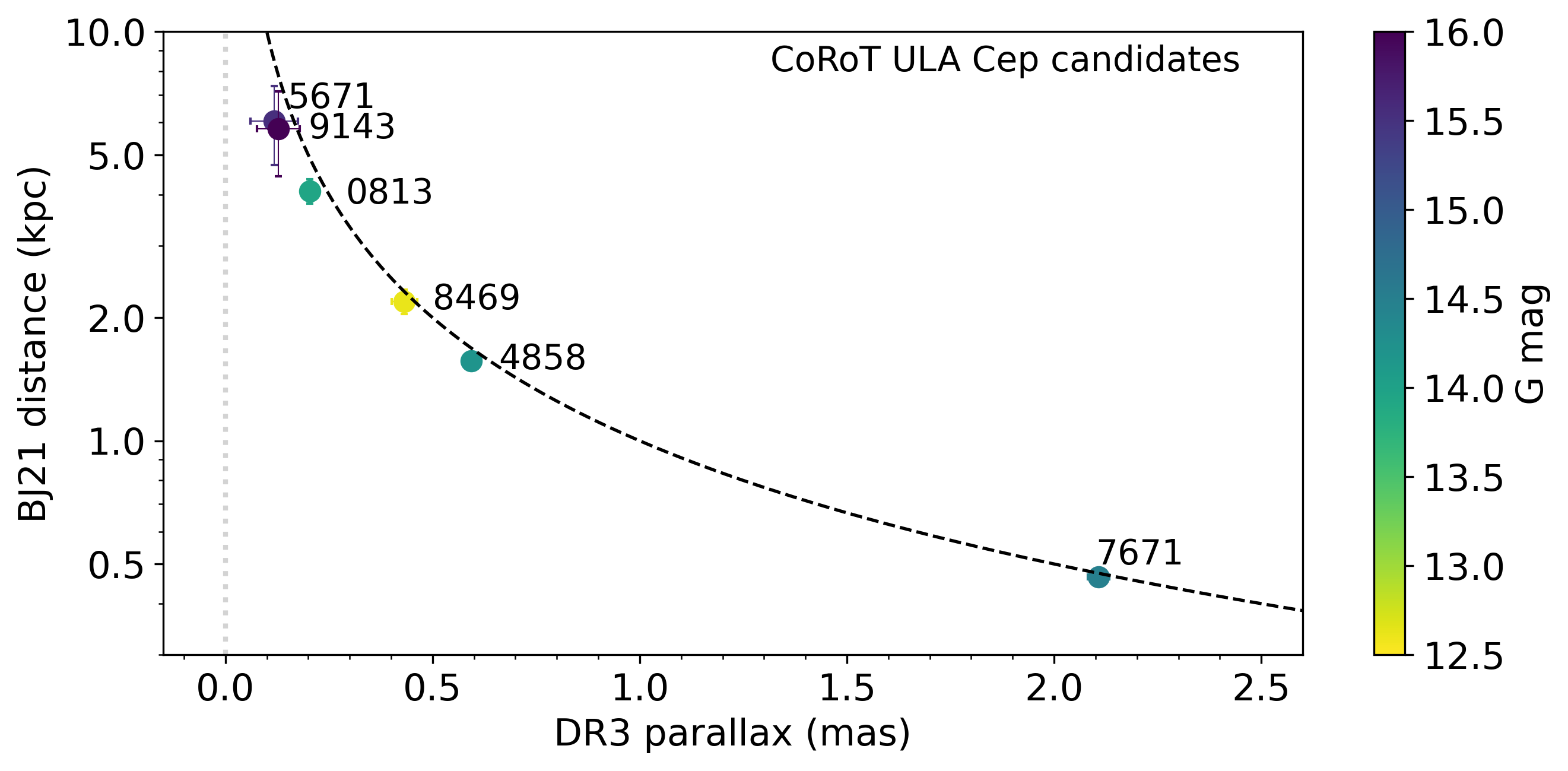}
    \caption{Gaia DR3  parallaxes ($\varpi$) versus geometric distances calculated by \cite{BailerJones2021} on a logarithmic scale for ULA Cepheid candidates with the measured and calculated error bars. It can be seen that the calculated distance for each Cepheid ULA candidate fits well with the  $\rho = 1 / \varpi$ parallax-distance inversion law (black dashed line). The apparent magnitudes in the $G$ band of each star are color coded.}
    \label{fig:plxoverULA}
\end{figure}  

\subsection{Gaia (E)DR3 data and absolute magnitudes}
\label{subsec:gaiadata}

   We found all six candidates in the main source catalog of the \textit{Gaia} DR3 database \citep{gaia-edr3-2021} based on the optical magnitude and position of each star.
   None of these stars are marked as variable stars (\texttt{phot\_variable\_flag} set to "NOT\_AVAILABLE"). 
   In order to collect all the relevant pieces of information about the ULA candidates automatically, we developed a routine in python to carry out the following steps in the case of (E)DR3 data. First, it uses the \texttt{astroquery} package of \citet{astroquery} to query the Gaia database for positions, parallaxes, as well as $G$, $G_\mathrm{RP}$, and $G_\mathrm{BP}$ magnitudes for each star. Next, the package queries the SIMBAD database for $V$ and 2MASS $J$, $H$, and $K$ magnitudes, and then it collects the calculated distances for each star from the catalog of \cite{BailerJones2021}. It subsequently calculates the extinction values for each star from the Bayestar19 combined dust maps of \cite{Greenetal2019}. Finally, it calculates the reddening-free absolute magnitudes in each Gaia, $V$, and 2MASS band. Table\,\ref{table:uladata} collects basic data from the Gaia archive: DR3 ID; $G$, $G_\mathrm{RP},$ and $G_\mathrm{BP}$ magnitudes; the distances calculated by \cite{BailerJones2021}; and the parallaxes for each star.

   \subsubsection{Additional data about specific variable star types}
   \label{sec:add_data}
   
  In order to specify the positions of the six candidates in the color--magnitude diagram (CMD), we collected additional data from the Gaia archive. First, we collected Cepheids from the variability table (\texttt{vari\_cepheid}) of the DR3 database for comparison.
   This table contains 15\,021 variable stars validated by the Specific Object Study (SOS) Cep\&RRL pipeline \citep{Clementinietal2016,Clementinietal2019}, of which more than 61\% are located in the Magellanic Clouds \citep{Ripepietal2022}.
   The database contains fundamental, first overtone, and multimode classical Cepheids, Type~II Cepheids, and anomalous Cepheids as well.
   However, based on the position of multimode, Type~II, and anomalous Cepheids in the CMD, these stars were not considered.
   The number of the remaining fundamental and first overtone Cepheids thus decreased to about 12\,800.

   We also collected data of slow pulsating B (SPB) stars and $\beta$ Cephei stars from the all-sky classifier of DR3 \citep[\texttt{vari\_classifier\_result} table,][]{rimoldini-2022}. 
   The steps of the data processing to achieve reddening-free absolute magnitudes are described in Section\,\ref{subsec:gaiadata}. 
   We also performed these steps for the Gaia Catalogue of Nearby Stars \citep{GCNS2021}, which contains data for more than 330\,000 stars within $100$\,pc.

   It is important to note that neither the spectral response for CoRoT AS \citep{Auvergneetal2009} and Gaia $G$ passbands \citep{GaiaG}, nor TESS \citep{TESS2015} and Gaia $G_\mathrm{RP}$ \citep{GaiaG} passbands are fully equivalent, which can affect the amplitudes and overall shapes of the light curves. 
   However, the transmission functions and their maxima are almost identical for the CoRoT AS and Gaia G passbands, so, in the present study, these two transmission functions are used together.

\section{Results}
\label{sec:res}

\subsection{Distances and absolute magnitudes}
\label{sec:distances}
   
   Gaia surveys the whole sky up to 20-22 magnitudes in the G band \citep{Gaiamission2016}. 
   Errors in parallax data can be significant for fainter stars since the accuracy of stellar positions primarily scales with the apparent brightness of stars.
   Thus, the accuracy of the parallax not only depends on the distance, but on the brightness of the object as well.
   Further, Gaia parallaxes are statistical data based on a set of measurements, and therefore negative parallaxes also occur in the database. 
   To handle this, the distances were not calculated as simple parallax inversions but were estimated via a probabilistic approach using a 3D model of the Milky Way as a prior.
   This way, very uncertain and negative parallaxes could be handled together with the rest of the data, although the resulting distances are understandably expected to still have large uncertainties \citep{BailerJones2015,BailerJones2021}.

   \begin{figure}
    \centering
    \includegraphics[width=\columnwidth]{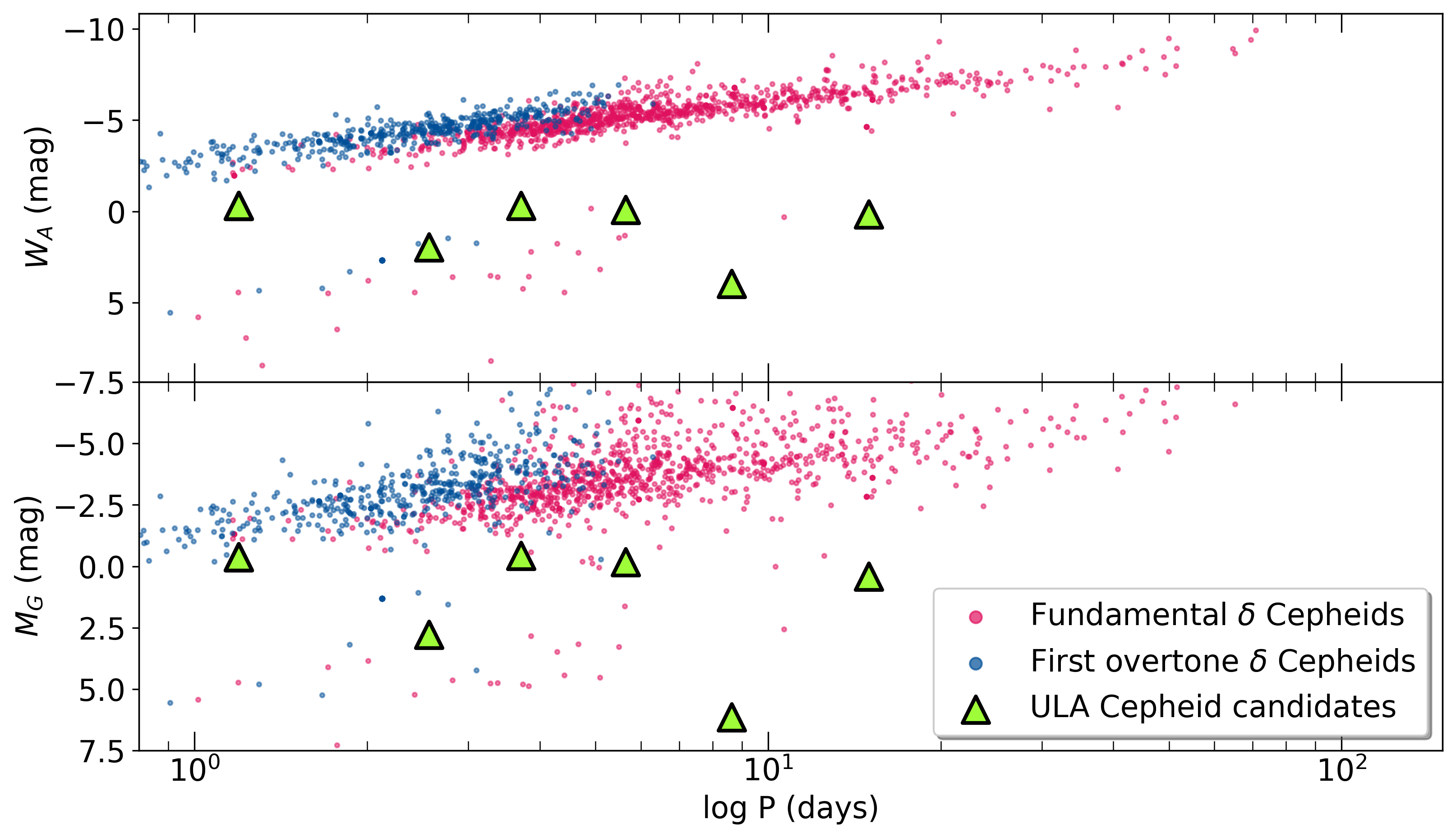}
    \caption{Period--Wesenheit (upper panel) and period--luminosity (lower panel) relation diagrams for the fundamental and first-overtone Cepheids, with the six ULA Cepheid candidates (green triangles).}
    \label{fig:pl-rel}
\end{figure}  

   Figure\,\ref{fig:plxoverULA} presents the accuracy of parallax values in the Gaia DR3 archive versus calculated distances by \cite{BailerJones2021}.
   It can be seen that both the calculated distance of \cite{BailerJones2021} and the parallax values of each star fit well with the $\rho=1/\varpi$ parallax--distance function. 
   The plot also shows that star 7671 is quite close to us at $\sim 500$ pc. 
   Together with its low apparent brightness, this already suggests it is a low-luminosity star rather than a Cepheid.
   
   We also checked the parallaxes of the variable star populations described in Section \ref{sec:add_data}, which we use for comparison purposes later. 
   Figures\,\ref{fig:plxover} and \ref{fig:plxover_cut} in the Appendix present the selected samples for each variable star type.
   These figures, similar to Figure\,\ref{fig:plxoverULA}, show DR3 parallax data versus calculated distances by \cite{BailerJones2021}, but for SPB and $\beta$ Cephei stars, as well as for fundamental and first overtone classical Cepheids. 
   We used a parallax-over-error threshold of $\mathbf{\varpi/\sigma_\varpi>5}$ to select stars for comparison, resulting in a large sample of variables per class that have relatively accurate distances estimated and thus have well-defined positions on a CMD.  
   The selected parallax-over-error threshold coincides with the limit used by \cite{Luri2021} to separate foreground Milky Way stars from Magellanic Clouds, which accounts for more than 61\% of the known Cepheids \citep{Ripepietal2022}. 
   Moreover, this limit also filters out stars in the Galactic bulge and stars with high galactic latitudes. 
   As a result, the number of the filtered classical fundamental and first overtone Cepheids was reduced to $\sim1550$.

\begin{figure}
    \centering
    \includegraphics[width=\columnwidth]{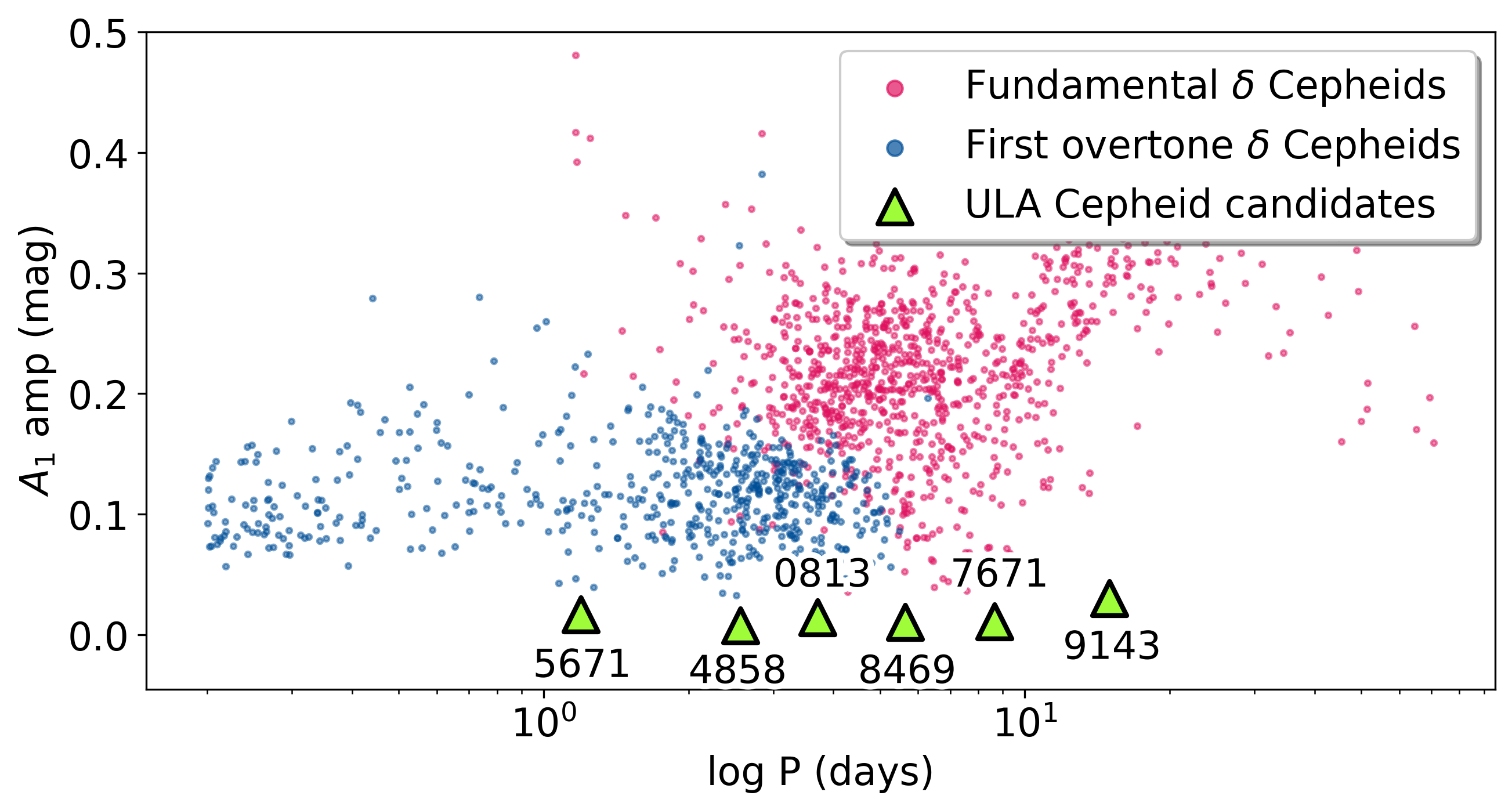}
    \caption{Period -- amplitude diagram for the fundamental (pink) and first overtone Cepheids (blue) from the Gaia DR3 archive \citep{Ripepietal2022}. The six ULA Cepheid candidates are marked with green triangles based on CoRoT data (see Table\,\ref{table:res_fourier}). 
    }
    \label{fig:per-ampl}
\end{figure}   

\begin{figure*}
    \centering
    \includegraphics[width=\textwidth]{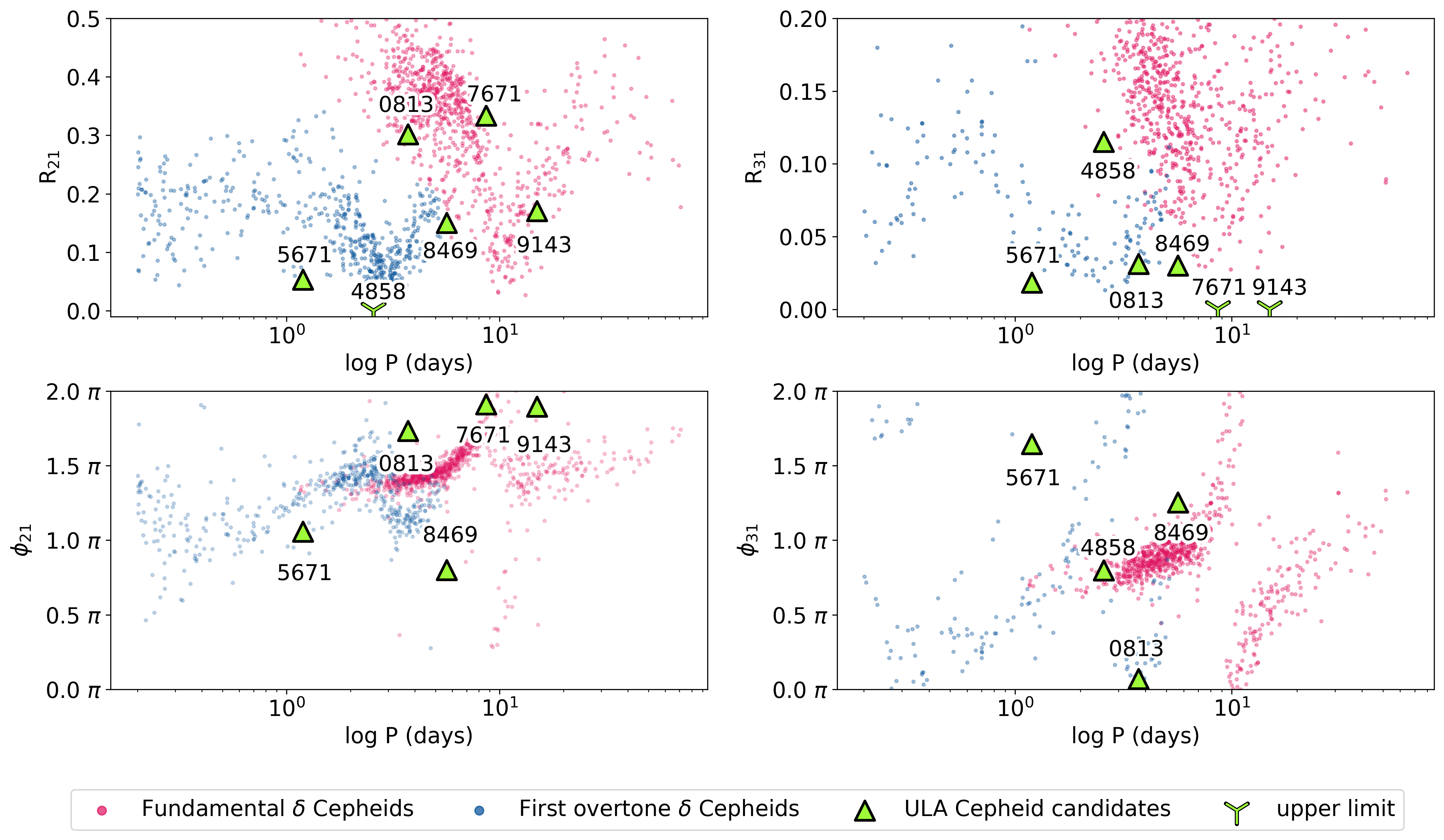}
    \caption{$R_{21}$, $R_{31}$, $\phi_{21}$, and $\phi_{31}$ relative Fourier parameters versus the logarithm of the period (in days) for the fundamental (red) and first overtone Cepheids (blue) from the Gaia DR3 archive \citep{Ripepietal2022}. The six ULA Cepheid candidates are marked with green triangles with values based on CoRoT data (see Table\,\ref{table:res_fourier}).}
    \label{fig:per-r21}
\end{figure*}

\subsection{Periods, amplitudes, and relative Fourier parameters}

   Table\,\ref{table:res_fourier} lists the Fourier parameters we calculated, as described in Section\,\ref{sec:fourierdesc}. 
   The periods calculated from the CoRoT observations match well with those calculated by \cite{Szaboetal2009}, with differences being mostly on the order of $0.01$\,days (compared with Table\,\ref{table:corot_data}). 
   However, the calculated period value of the star identified as 9143 differs from that listed in Table\,\ref{table:corot_data} by $\sim 0.15$\,days. 
   We also found the harmonics indicated in Table\,\ref{table:corot_data} for each star (e.g., $2f_0$ and $3f_0$ for star 5671).
   In addition, we found the third harmonic of the dominant frequency in stars 0813 and 8469.
   Similarly, the amplitude of the light variation presented here agrees well with the previous data on the millimagnitude-level. 
   Thus, the ULA brightness variation of each star in CoRoT data is confirmed in this study as well.
   
   In the case of TESS data, we found that period values and amplitudes for stars 5671, 4858, 0813, and 8469 match well with CoRoT data in both sectors, for example, the difference of periods between TESS and CoRoT data is on the order of $0.001 - 0.01$ days.
   On the other hand, it can be seen that the light curves for stars 7671 and 9143 are either contaminated or too noisy to identify an obvious periodic brightness variation in either sectors. 
   Further, the period of 9143 is very close to the orbital period of TESS, making the identification of its variability even more difficult.

   As a first step, the calculated period values (based on CoRoT data) were used to locate the position of the six stars compared to Leavitt's law, both on the period versus\ reddening-free Wesenheit (PW$_A$), and the period--luminosity (PL) diagrams (see Figure\,\ref{fig:pl-rel}), where $W_\mathrm{A} = M_\mathrm{G} - 1.9 (M_\mathrm{BP} - M_\mathrm{RP})$ \cite[see, e.g.,][]{Ripepietal2019}. 
   In the case of high-overtone or ``strange mode" Cepheids, the positions of the stars would be above the first-overtone and fundamental-mode ones on the PL diagram.
   
   Additionally, Figures \ref{fig:per-ampl}  and \ref{fig:per-r21} present the $A_1$ Fourier amplitudes (the \texttt{fund\_freq1\_harmonic\_ampl\_g} values in DR3) versus the pulsation periods, and the relative Fourier parameters ($R_{21}$, $R_{31}$, $\phi_{21}$, and $\phi_{31}$) versus the pulsation periods for the Gaia DR3 fundamental and first overtone Cepheids (red and blue points, from the \texttt{vari\_cepheid} table) and the six ULA candidates (green triangles), respectively, based on the CoRoT light curves. 
   In those cases, when the amplitude of either the second or third harmonic was not detected with an amplitude exceeding a signal-to-noise ratio of three compared to the local noise level of the frequency spectrum, we used the detection threshold to calculate upper limits for $R_{21}$ and $R_{31}$.
   These upper limits are marked with green tri-down symbols in Figure\,\ref{fig:per-r21}. 
   Clearly, although the period values fall within the range of classical Cepheids, the amplitudes of the brightness variations are lower than that of the known Cepheids in all cases. 
   Furthermore, Figure \ref{fig:per-r21} shows that none of the stars can be clearly classified either as fundamental or as first-overtone Cepheids. 
   As an example, the $R_{21}$ value of star 0813 falls into the range of fundamental Cepheids; however, its $R_{31}$ value is more similar to that of fundamental mode Cepheids. 
   Moreover, its $\phi_{21}$ and $\phi_{31}$ values fall outside the ranges of both fundamental and first overtone Cepheids. 
   In general, none of the six stars match all four Fourier parameters of either the fundamental mode or the first overtone Cepheids simultaneously.
   Thus, we can conclude that none of the stars are fundamental, or first-overtone Cepheids. 
   Although, it cannot be ruled out that they are strange mode pulsators. 
   Therefore, we identified their location in the CMD in the next section.

\begin{figure*}
    \centering
    \includegraphics[width=\textwidth]{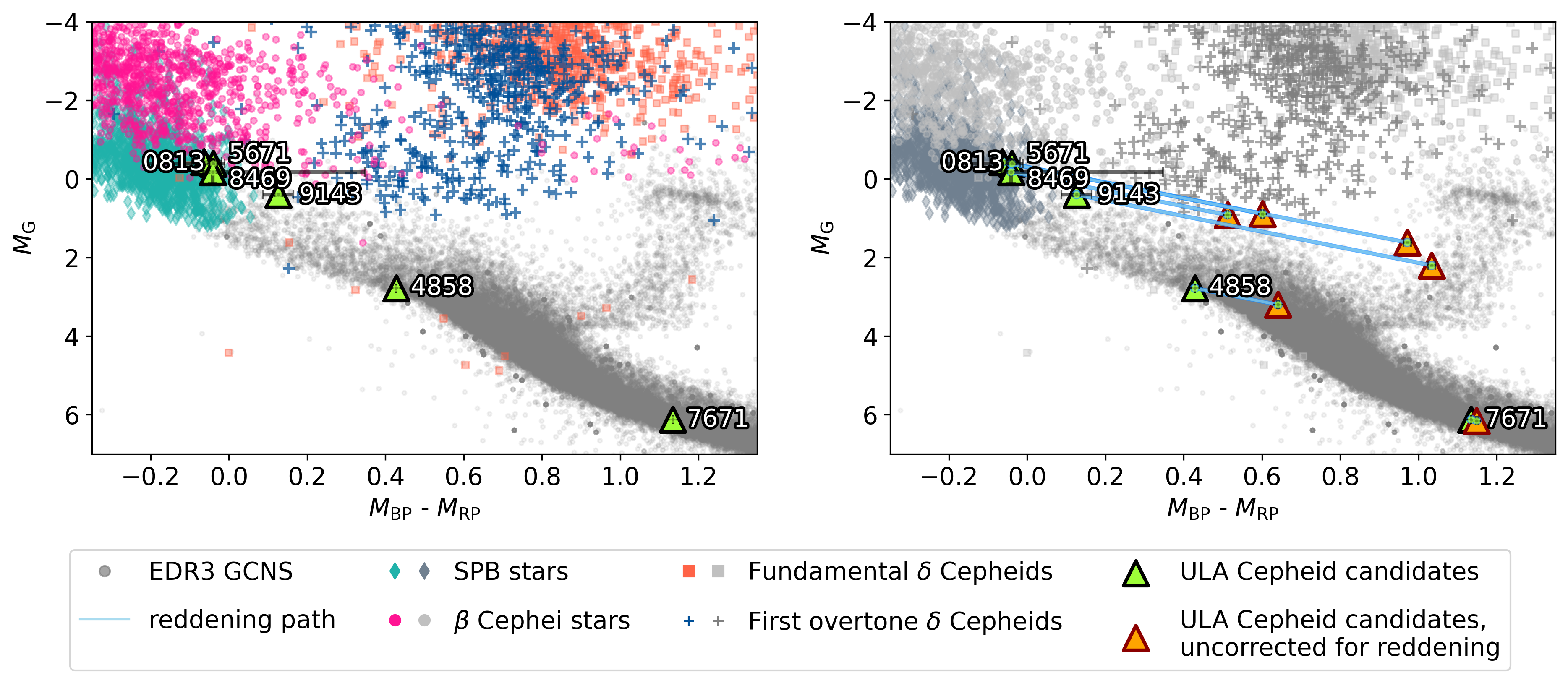}
    \caption{Color--magnitude diagrams (CMDs) of Gaia $M_\mathrm{G}$ versus the reddening-free $(M_\mathrm{BP}-M_\mathrm{RP})$ color indices. In the left panel, stars from the Gaia Catalogue of Nearby Stars of \cite{GCNS2021} are marked with gray dots.\ In addition, SPB and $\beta$ Cephei, as well as fundamental and first overtone Cepeheid stars from DR3 are marked with green diamonds, pink dots, orange squares and blue crosses, respectively. The six ULA Cepheid candidates are marked with green plus orange triangles (in the right panel). The right panel is the same as the left panel, but with background stars shown in gray scale marked with the same symbols. Absolute magnitudes were calculated from the distances of \citet{BailerJones2021}. The reddening vectors are shown with a solid blue line.}
    \label{fig:camd}
\end{figure*}

\subsection{The color--magnitude diagrams}
\label{sect:gaiacmd}
   Using the Gaia EDR3 distances of \cite{BailerJones2021} and the calculated dereddened, absolute magnitudes in Gaia $M_\mathrm{G}$, $M_\mathrm{RP}$, and $M_\mathrm{BP}$ and infrared 2MASS $M_J$, $M_H,$ and $M_K$ bands, we identified the positions of the six ULA Cepheid candidates in the CMDs.
   In the left panel of Figure~\ref{fig:camd}, gray dots denote stars within 100~pc from the Gaia Catalogue of Nearby Stars, compiled by \citet{GCNS2021}, while green diamonds and pink dots are SPB and $\beta$ Cephei stars from the Gaia DR3 variability classification table (\texttt{vari\_classifier\_result}). CMDs in the infrared based on 2MASS $J$, $H$, and $K$ bands are shown in Figure\,\ref{fig:camdIR} in the Appendix. The markings for both of the background stars and the ULA Cepheid candidates are the same as those shown in Figure\,\ref{fig:camd}.
   
   The left panel of Figure\,\ref{fig:camd} presents the CMD in Gaia $M_{\rm G}$ brightness versus the reddening-free $M_\mathrm{BP}-M_\mathrm{BP}$ color index.
   Fundamental and first overtone Cepheids filtered by the $20\%$ statistical error limit in parallax are marked by orange squares and blue crosses based on the work of \cite{Ripepietal2022}. 
   For comparison, in the right panel of Figure\,\ref{fig:camd}, we also show the CMDs with the absolute magnitudes of each ULA Cepheid candidate calculated from the \citet{BailerJones2021} distances, but neglecting the correction for interstellar absorption and reddening (orange triangles). 
   The reddening vectors are marked with blue lines for each star. 
   It can be seen that neglecting the correction for reddening, similarly, in the case of reddening-free absolute distances, none of the stars fall within the range bounded by the classical Cepheids (marked with gray triangles and crosses).

\subsubsection{Stars 4858 and 7671}
    
   Based on their position on the CMDs, stars 4858 and 7671 appear to be main-sequence A and K stars.
   The former range is populated by $\delta$ Scuti, SX Phoenicis, $\gamma$ Doradus, and rapidly oscillating Am or Ap (roAm and roAp) stars \citep[see, e.g.,][and the references therein]{Skarkaetal2022} that show low amplitude light variations (see Figure\,\ref{fig:camd}). 
   
   However, in the case of, for example, $\delta$ Scutis, the pulsation period and the amplitude of light variation are on the order of a few hours and a few tenths of magnitudes \citep[see, e.g.,][]{Breger2000}, respectively (compared to a few days and a few millimagnitudes for our targets). 
   SX Phoenicis stars are blue stragglers in globular clusters in the galactic halo \citep[see, e.g.,][]{Jeonetal2004}, which contradicts the fact that all six ULA Cepheid candidates are found in the anticenter direction of the Galaxy.
   Going further, $\gamma$ Doradus stars are multi-periodic $g$-mode pulsators with periods on the order of a few hours to days \citep[see, e.g.,][]{Kayeetal1999}. 

   Although the amplitude of the light variation of rapidly oscillating peculiar stars (roAm and roAp stars) is on the order of a few millimagnitudes, the periods of their light variations are on the order of a few minutes, or hours. 
   In contrast, main-sequence K stars do not show coherent pulsations at all. 
   Thus, we can conclude that neither 4858 nor 7671 belong to the abovementioned periodic main-sequence variable star types. 
   These stars are rather likely to be A- and K-type spotted rotational variables.

\subsubsection{Stars 0813, 5671, 8469, and 9143}   
   
   Stars 0813, 5671, 8469, and 9143 are located close to the blue tip of the main sequence in the CMDs, where slowly pulsating B stars reside \citep[as well as a few known $\beta$ Cephei stars, see, e.g.,][]{DR2puls}.
   SPB stars are multi-periodic $g$-mode variables, and thus their light curves are less regular in time \citep[see, e.g.,][]{GaiaMSpuls2022}. 
   This contradicts the basic ULA Cepheid candidate selection requirement of only one dominant frequency and its possible harmonics being present \citep[see][]{Szaboetal2009}. 
   Moreover, the amplitudes of the light variation of these stars are, on average, approximately ten times larger than that of the four candidates \citep[see, e.g.,][]{Waelkens1991,Waelkens1996,Pedersenetal2021}. Thus, we can exclude that these stars are SPB stars.
   
   Furthermore, it is also doubtful that these stars are $\beta$~Cephei type variables. $\beta$~Cephei variables are hot blue stars, with typical pulsation periods and amplitudes on the order of a few hours and from $\sim 0.01$ to $\sim 0.1$ magnitudes, respectively \citep[see, e.g.,][and references therein]{SterkenandJerzykiewicz1993,GaiaMSpuls2022}.
   Since the periods of the four stars are longer than a few hours, we can conclude that these stars are not $\beta$~Cephei variables either. 
   
   This leads to the conclusion that these are likely not pulsating variable stars. 
   Rather, they are rotation-induced variables, for example, $\alpha^2$ Canum Venaticorum stars that are typically A- or B-type main-sequence stars with a brightness variability on the order of 0.01 to 0.1 magnitudes and light variation periods falling into the range of 0.5 to 160 days \citep[see, e.g.,][and the references therein]{Cunhaetal2019,rimoldini-2022}.

\subsection{O--C diagrams}

\begin{figure}
    \centering
    \includegraphics[width=\columnwidth]{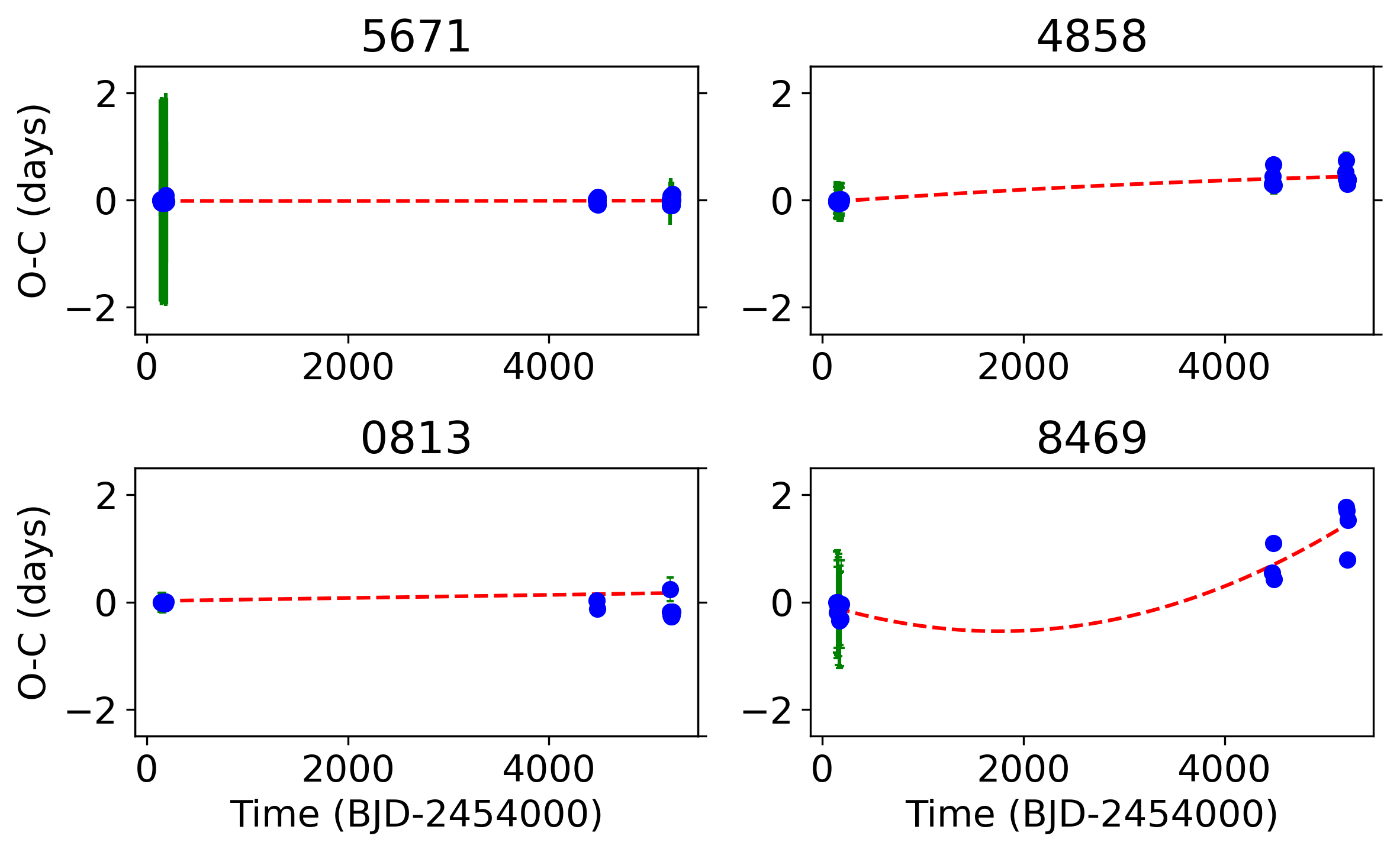}
    \caption{Observed--calculated (O--C) diagrams for four out of the six ULA Cepheid candidates of \cite{Szaboetal2009} based on CoRoT IRa01 and TESS data from sectors 06 and 33.}
    \label{fig:omc}
\end{figure} 

   We calculated the long-term O--C changes from the timings of the light-curve maxima for the four stars where we were able to determine the light variation periods from the TESS observations (stars 5671, 4858, 0813, and 8469, see Table\,\ref{table:res_fourier}). 
   The most noticeable change can be seen in the case of star 8469 (bottom right panel of Figure\,\ref{fig:omc}).
   Although, due to the nonuniform temporal distribution of the measurements, we do not know the exact rate of the change; it is clearly visible that even between the measurements of the two TESS sectors there are differences in the O-C data.
   This indicates a possible period change in the variation on timescales on the order of 10 years.
  
   No phase deviation was found between the CoRoT and TESS measurements in the case of star 5671. 
   Thus, we can state that the period of the light variation of this star has not changed on a 10-year timescale. 
   Contrary to the former two stars, 4858 and 0813 show some variation in phase, although those variations might come from small inaccuracies in the period determination accumulating over the time span between the CoRoT and TESS observations.

\section{Discussion and conclusions}
\label{sec:concl}

   In this paper, we have investigated six stars proposed to be ULA Cepheid candidates in the Galactic plane based on CoRoT observations analyzed by \citet{Szaboetal2009}.
   Additionally, we collected 2MASS, Gaia DR3, and TESS data for each star.
   None of the six stars were marked as variable in the Gaia DR3 archive, and thus photometric measurements were not available for any of them. 
   However, we were able to identify low-amplitude variations in the TESS data (in sectors 6 and 33) in four out of six candidates.

   Data processing of the TESS FFIs was performed with the \texttt{FITSH} software package of \cite{Pál2012}. 
   Absolute magnitudes were calculated by a python tool built by our team, based on Gaia EDR3 geometric distances of \cite{BailerJones2021}, from which we subtracted the reddening based on the Bayestar19 dust map of \cite{Greenetal2019}. 
   Based on their absolute magnitudes, we placed the six candidates on the $M_\mathrm{G}$--$(M_\mathrm{BP}-M_\mathrm{RP})$ CMDs and compared their positions to some known pulsating variables, for example, fundamental and first overtone Cepheids, slowly pulsating B (SPB) stars, and $\beta$ Cephei stars (see, e.g., in the left panel of Figure\,\ref{fig:camd}). 
   For comparison, we also calculated their positions in the CMD neglecting absorption correction to show the strong reddening some of them experience (see right panel of Figure\,\ref{fig:camd}).
   Infrared CMDs are shown in Figure\,\ref{fig:camdIR} (i.e., $M_\mathrm{K}$ versus $(M_\mathrm{J}-M_\mathrm{K})$ and $M_\mathrm{K}$ versus $(M_\mathrm{H}-M_\mathrm{K})$  CMDs based on 2MASS data). Markings of stars in Figure\,\ref{fig:camdIR} are the same as in Figure\,\ref{fig:camd}.

   In addition, we recalculated the periods and amplitudes of the light variation and the relative Fourier parameters for each star, based on the CoRoT data, with another python tool built by our team and the \texttt{PERIOD04} package of \citet{LenzandBreger2005}. 
   We could find all the harmonics of the dominant frequency that had been determined earlier. We also found the third harmonic of the dominant frequency in stars 0813 and 8469.
   Periods and Fourier parameters were calculated from the TESS data as well, and the extent of the long-term phase and period change was established for four out of six stars. 
   
   Our findings can be summarized as follows:

   \begin{enumerate}
       \item Pulsation period values and Fourier parameters presented in this work, based on CoRoT observations, agree well with the values published by \cite{Szaboetal2009}. 
       We have demonstrated that period data and the amplitude of the variation calculated from CoRoT and TESS data match well. 
       For the period data, the difference, calculated from the measurement data for the space telescopes, is on the order of $0.001 - 0.01$ days, while the difference for the amplitude of light variation is on the order  of millimagnitudes.
       \item The stars fall below the period--luminosity and period--Wesenheit relations of classical Cepheids, whereas strange-mode stars pulsating in high overtones are expected to be above them.
       \item We conclude that the $R_{21}$, $R_{31}$, $\phi_{21}$, and $\phi_{31}$ values do not fall simultaneously into the range of that of the classical Cepheids for any of the six stars.
       \item Furthermore, we find that none of the six ULA Cepheid candidates are located in the region populated by the fundamental or first overtone Cepheids in the CMDs.
       \item Stars 4858 and 7671 are located in the range of A- and K-type main--sequence stars. 
       However, pulsation periods of main-sequence pulsating variables are mainly on the order of a few hours, and thus we conclude that these stars are not main-sequence pulsators, but rather they are likely  rotational spotted variables.
       \item Stars 0813, 5671, 8469, and 9143 are located in the region populated by slowly pulsating B and $\beta$ Cephei stars. 
       As SPB stars are multiperiodic $g$-mode pulsators and the period of light variation of $\beta$ Cepheis is on the order of a few hours, we conclude that these four are not likely to be pulsating variable stars either. 
       Rather, they are also rotation-induced variables, such as $\alpha^2$~Canum Venaticorum stars.
   \end{enumerate}

   It is important to point out that additional measurements are required for the six candidates presented here to be able to determine their variability types precisely. 
   However, we can conclude that none of the six ULA Cepheid candidates are located in the loci of  the classical Cepheids independent of bandpasses and regardless of whether we corrected for reddening or not. Rather, these stars are more likely to be rotation-induced K-, A-, or B-type main-sequence variable stars (spotted and $\alpha^2$~Canum Venaticorum stars), rather than pulsating variables.

   Finally, it is also important to point out that our knowledge about ULA Cepheids is currently very limited, and these targets did not advance our understanding either. 
   The exact mechanism of the formation of ULA, or even strange mode pulsation is not well known, nor is the exact location of these stars in the Hertzsprung-Russell diagram and the CMDs. 
   Therefore, more detailed evolutionary and pulsation modeling will be extremely important, as will be targeted searches in astronomical observations, to test the existence of ULA Cepheid stars more conclusively.

\begin{acknowledgements}
This project has been supported by the KKP-137523 'SeismoLab' Élvonal grant of the Hungarian Research, Development and Innovation Office (NKFIH), the Lendület Program of the Hungarian Academy of Sciences under project No.\ LP2018-7/2022, and the Hungarian OTKA Grant No.\ 119993. This paper includes data collected by the TESS mission, which is publicly available from the Mikulski Archive for Space Telescopes (MAST). Funding for the TESS mission is provided by NASA's Science Mission directorate. This work has made use of data from the European Space Agency (ESA) mission Gaia (\url{https://www.cosmos.esa.int/gaia}), processed by the Gaia Data Processing and Analysis Consortium (DPAC, \url{https://www.cosmos.esa.int/web/gaia/dpac/consortium}). Funding for the DPAC has been provided by national institutions, in particular, the institutions participating in the Gaia Multilateral Agreement. This research has made use of the Spanish Virtual Observatory (\url{https://svo.cab.inta-csic.es}) project funded by MCIN/AEI/10.13039/501100011033/ through grant PID2020-112949GB-I00, of NASA’s Astrophysics Data System Bibliographic Services, and of the SIMBAD database and "Aladin sky atlas", operated at CDS, Strasbourg, France.

\end{acknowledgements}

%
%

\bibliographystyle{aa} 
\bibliography{references} 

\begin{thebibliography}{52}
\expandafter\ifx\csname natexlab\endcsname\relax\def\natexlab#1{#1}\fi

\bibitem[{{Auvergne} {et~al.}(2009){Auvergne}, {Bodin}, {Boisnard}, {Buey},
  {Chaintreuil}, {Epstein}, {Jouret}, {Lam-Trong}, {Levacher}, {Magnan},
  {Perez}, {Plasson}, {Plesseria}, {Peter}, {Steller}, {Tiph{\`e}ne}, {Baglin},
  {Agogu{\'e}}, {Appourchaux}, {Barbet}, {Beaufort}, {Bellenger}, {Berlin},
  {Bernardi}, {Blouin}, {Boumier}, {Bonneau}, {Briet}, {Butler}, {Cautain},
  {Chiavassa}, {Costes}, {Cuvilho}, {Cunha-Parro}, {de Oliveira Fialho},
  {Decaudin}, {Defise}, {Djalal}, {Docclo}, {Drummond}, {Dupuis}, {Exil},
  {Faur{\'e}}, {Gaboriaud}, {Gamet}, {Gavalda}, {Grolleau}, {Gueguen},
  {Guivarc'h}, {Guterman}, {Hasiba}, {Huntzinger}, {Hustaix}, {Imbert},
  {Jeanville}, {Johlander}, {Jorda}, {Journoud}, {Karioty}, {Kerjean},
  {Lafond}, {Lapeyrere}, {Landiech}, {Larqu{\'e}}, {Laudet}, {Le Merrer},
  {Leporati}, {Leruyet}, {Levieuge}, {Llebaria}, {Martin}, {Mazy}, {Mesnager},
  {Michel}, {Moalic}, {Monjoin}, {Naudet}, {Neukirchner}, {Nguyen-Kim},
  {Ollivier}, {Orcesi}, {Ottacher}, {Oulali}, {Parisot}, {Perruchot},
  {Piacentino}, {Pinheiro da Silva}, {Platzer}, {Pontet}, {Pradines},
  {Quentin}, {Rohbeck}, {Rolland}, {Rollenhagen}, {Romagnan}, {Russ}, {Samadi},
  {Schmidt}, {Schwartz}, {Sebbag}, {Smit}, {Sunter}, {Tello}, {Toulouse},
  {Ulmer}, {Vandermarcq}, {Vergnault}, {Wallner}, {Waultier}, \&
  {Zanatta}}]{Auvergneetal2009}
{Auvergne}, M., {Bodin}, P., {Boisnard}, L., {et~al.} 2009, \aap, 506, 411

\bibitem[{{Baglin} {et~al.}(2006){Baglin}, {Auvergne}, {Barge}, {Deleuil},
  {Catala}, {Michel}, {Weiss}, \& {COROT Team}}]{corot-2006}
{Baglin}, A., {Auvergne}, M., {Barge}, P., {et~al.} 2006, in ESA Special
  Publication, Vol. 1306, The CoRoT Mission Pre-Launch Status - Stellar
  Seismology and Planet Finding, ed. M.~{Fridlund}, A.~{Baglin}, J.~{Lochard},
  \& L.~{Conroy}, 33

\bibitem[{{Bailer-Jones}(2015)}]{BailerJones2015}
{Bailer-Jones}, C. A.~L. 2015, \pasp, 127, 994

\bibitem[{{Bailer-Jones} {et~al.}(2021){Bailer-Jones}, {Rybizki}, {Fouesneau},
  {Demleitner}, \& {Andrae}}]{BailerJones2021}
{Bailer-Jones}, C.~A.~L., {Rybizki}, J., {Fouesneau}, M., {Demleitner}, M., \&
  {Andrae}, R. 2021, \aj, 161, 147

\bibitem[{{Borkovits} {et~al.}(2022){Borkovits}, {Mitnyan}, {Rappaport},
  {Pribulla}, {Powell}, {Kostov}, {B{\'\i}r{\'o}}, {Cs{\'a}nyi}, {Garai},
  {Gary}, {Kaye}, {Kom{\v{z}}{\'\i}k}, {Terentev}, {Omohundro}, {Gagliano},
  {Jacobs}, {Kristiansen}, {LaCourse}, {Schwengeler}, {Czavalinga}, {Seli},
  {Huang}, {P{\'a}l}, {Vanderburg}, {Rodriguez}, \& {Stevens}}]{borkovits-2022}
{Borkovits}, T., {Mitnyan}, T., {Rappaport}, S.~A., {et~al.} 2022, \mnras, 510,
  1352

\bibitem[{{Breger}(2000)}]{Breger2000}
{Breger}, M. 2000, in Astronomical Society of the Pacific Conference Series,
  Vol. 210, Delta Scuti and Related Stars, ed. M.~{Breger} \& M.~{Montgomery},
  3

\bibitem[{{Breger} {et~al.}(1999){Breger}, {Handler}, {Garrido}, {Audard},
  {Zima}, {Papar{\'o}}, {Beichbuchner}, {Li}, {Jiang}, {Liu}, {Zhou}, {Pikall},
  {Stankov}, {Guzik}, {Sperl}, {Krzesinski}, {Ogloza}, {Pajdosz}, {Zola},
  {Thomassen}, {Solheim}, {Serkowitsch}, {Reegen}, {Rumpf}, {Schmalwieser}, \&
  {Montgomery}}]{Breger1999}
{Breger}, M., {Handler}, G., {Garrido}, R., {et~al.} 1999, \aap, 349, 225

\bibitem[{{Buchler} \& {Koll{\'a}th}(2001)}]{BuchlerandKollath2001}
{Buchler}, J.~R. \& {Koll{\'a}th}, Z. 2001, \apj, 555, 961

\bibitem[{{Buchler} {et~al.}(2005){Buchler}, {Wood}, {Keller}, \&
  {Soszy{\'n}ski}}]{Buchleretal2005}
{Buchler}, J.~R., {Wood}, P.~R., {Keller}, S., \& {Soszy{\'n}ski}, I. 2005,
  \apjl, 631, L151

\bibitem[{{Buchler} {et~al.}(2009){Buchler}, {Wood}, \&
  {Soszy{\'n}ski}}]{Buchleretal2009}
{Buchler}, J.~R., {Wood}, P.~R., \& {Soszy{\'n}ski}, I. 2009, \apj, 698, 944

\bibitem[{{Buchler} {et~al.}(1997){Buchler}, {Yecko}, \&
  {Kollath}}]{Buchleretal1997}
{Buchler}, J.~R., {Yecko}, P.~A., \& {Kollath}, Z. 1997, \aap, 326, 669

\bibitem[{{Clementini} {et~al.}(2016){Clementini}, {Ripepi}, {Leccia},
  {Mowlavi}, {Lecoeur-Taibi}, {Marconi}, {Szabados}, {Eyer}, {Guy},
  {Rimoldini}, {Jevardat de Fombelle}, {Holl}, {Busso}, {Charnas}, {Cuypers},
  {De Angeli}, {De Ridder}, {Debosscher}, {Evans}, {Klagyivik}, {Musella},
  {Nienartowicz}, {Ord{\'o}{\~n}ez}, {Regibo}, {Riello}, {Sarro}, \&
  {S{\"u}veges}}]{Clementinietal2016}
{Clementini}, G., {Ripepi}, V., {Leccia}, S., {et~al.} 2016, \aap, 595, A133

\bibitem[{{Clementini} {et~al.}(2019){Clementini}, {Ripepi}, {Molinaro},
  {Garofalo}, {Muraveva}, {Rimoldini}, {Guy}, {Jevardat de Fombelle},
  {Nienartowicz}, {Marchal}, {Audard}, {Holl}, {Leccia}, {Marconi}, {Musella},
  {Mowlavi}, {Lecoeur-Taibi}, {Eyer}, {De Ridder}, {Regibo}, {Sarro},
  {Szabados}, {Evans}, \& {Riello}}]{Clementinietal2019}
{Clementini}, G., {Ripepi}, V., {Molinaro}, R., {et~al.} 2019, \aap, 622, A60

\bibitem[{{CoRoT Team}(2016{\natexlab{a}})}]{CorotBook2016}
{CoRoT Team}. 2016{\natexlab{a}}, {The CoRoT Legacy Book: The adventure of the
  ultra high precision photometry from space, by the CoRot Team} (EDP Sciences)

\bibitem[{{CoRoT Team}(2016{\natexlab{b}})}]{CorotFaint}
{CoRoT Team}. 2016{\natexlab{b}}, VizieR Online Data Catalog, B/corot

\bibitem[{{Cox} {et~al.}(1980){Cox}, {Wheeler}, {Hansen}, {King}, {Cox}, \&
  {Hodson}}]{Coxetal1980}
{Cox}, J.~P., {Wheeler}, J.~C., {Hansen}, C.~J., {et~al.} 1980, \ssr, 27, 529

\bibitem[{{Cunha} {et~al.}(2019){Cunha}, {Antoci}, {Holdsworth}, {Kurtz},
  {Balona}, {Bogn{\'a}r}, {Bowman}, {Guo}, {Ko{\l}aczek-Szyma{\'n}ski},
  {Lares-Martiz}, {Paunzen}, {Skarka}, {Smalley}, {S{\'o}dor}, {Kochukhov},
  {Pepper}, {Richey-Yowell}, {Ricker}, {Seager}, {Buzasi}, {Fox-Machado},
  {Hasanzadeh}, {Niemczura}, {Quitral-Manosalva}, {Monteiro}, {Stateva}, {De
  Cat}, {Garc{\'\i}a Hern{\'a}ndez}, {Ghasemi}, {Handler}, {Hey}, {Matthews},
  {Nemec}, {Pascual-Granado}, {Safari}, {Su{\'a}rez}, {Szab{\'o}}, {Tkachenko},
  \& {Weiss}}]{Cunhaetal2019}
{Cunha}, M.~S., {Antoci}, V., {Holdsworth}, D.~L., {et~al.} 2019, \mnras, 487,
  3523

\bibitem[{{Cutri} {et~al.}(2003){Cutri}, {Skrutskie}, {van Dyk}, {Beichman},
  {Carpenter}, {Chester}, {Cambresy}, {Evans}, {Fowler}, {Gizis}, {Howard},
  {Huchra}, {Jarrett}, {Kopan}, {Kirkpatrick}, {Light}, {Marsh}, {McCallon},
  {Schneider}, {Stiening}, {Sykes}, {Weinberg}, {Wheaton}, {Wheelock}, \&
  {Zacarias}}]{Cutrietal2003}
{Cutri}, R.~M., {Skrutskie}, M.~F., {van Dyk}, S., {et~al.} 2003, VizieR Online
  Data Catalog, II/246

\bibitem[{{Czavalinga} {et~al.}(2023){Czavalinga}, {Mitnyan}, {Rappaport},
  {Borkovits}, {Gagliano}, {Omohundro}, {Kristiansen}, \&
  {P{\'a}l}}]{czavalinga-2022}
{Czavalinga}, D.~R., {Mitnyan}, T., {Rappaport}, S.~A., {et~al.} 2023, \aap,
  670, A75

\bibitem[{{Gaia Collaboration} {et~al.}(2021{\natexlab{a}}){Gaia
  Collaboration}, {Brown}, {Vallenari}, {Prusti}, {de Bruijne}, {Creevey},
  {Evans}, {Eyer}, {Hutton}, {Jansen}, \& {Jordi}}]{gaia-edr3-2021}
{Gaia Collaboration}, {Brown}, A.~G., {Vallenari}, A., {et~al.}
  2021{\natexlab{a}}, \aap, 649, A1

\bibitem[{{Gaia Collaboration} {et~al.}(2022){Gaia Collaboration}, {De Ridder},
  {Ripepi}, {Aerts}, {Palaversa}, {Eyer}, {Holl}, {Audard}, {Rimoldini},
  {Brown}, {Vallenari}, {Prusti}, {de Bruijne}, {Arenou}, {Babusiaux},
  {Biermann}, {Creevey}, {Ducourant}, {Evans}, {Guerra}, {Hutton}, {Jordi},
  {Klioner}, {Lammers}, {Lindegren}, {Luri}, {Mignard}, {Panem}, {Pourbaix},
  {Randich}, {Sartoretti}, {Soubiran}, {Tanga}, {Walton}, {Bailer-Jones},
  {Bastian}, {Drimmel}, {Jansen}, {Katz}, {Lattanzi}, {van Leeuwen}, {Bakker},
  {Cacciari}, {Casta{\~n}eda}, {De Angeli}, {Fabricius}, {Fouesneau},
  {Fr{\'e}mat}, {Galluccio}, {Guerrier}, {Heiter}, {Masana}, {Messineo},
  {Mowlavi}, {Nicolas}, {Nienartowicz}, {Pailler}, {Panuzzo}, {Riclet}, {Roux},
  {Seabroke}, {Sordo}, {Th{\'e}venin}, {Gracia-Abril}, {Portell}, {Teyssier},
  {Altmann}, {Andrae}, {Bellas-Velidis}, {Benson}, {Berthier}, {Blomme},
  {Burgess}, {Busonero}, {Busso}, {C{\'a}novas}, {Carry}, {Cellino}, {Cheek},
  {Clementini}, {Damerdji}, {Davidson}, {de Teodoro}, {Nu{\~n}ez Campos},
  {Delchambre}, {Dell'Oro}, {Esquej}, {Fern{\'a}ndez-Hern{\'a}ndez}, {Fraile},
  {Garabato}, {Garc{\'\i}a-Lario}, {Gosset}, {Haigron}, {Halbwachs}, {Hambly},
  {Harrison}, {Hern{\'a}ndez}, {Hestroffer}, {Hilger}, {Hodgkin}, {Jan{\ss}en},
  {Jevardat de Fombelle}, {Jordan}, {Krone-Martins}, {Lanzafame},
  {L{\"o}ffler}, {Marchal}, {Marrese}, {Moitinho}, {Muinonen}, {Osborne},
  {Pancino}, {Pauwels}, {Recio-Blanco}, {Reyl{\'e}}, {Riello}, {Roegiers},
  {Rybizki}, {Sarro}, {Siopis}, {Smith}, {Sozzetti}, {Utrilla}, {van Leeuwen},
  {Abbas}, {{\'A}brah{\'a}m}, {Abreu Aramburu}, {Aguado}, {Ajaj},
  {Aldea-Montero}, {Altavilla}, {{\'A}lvarez}, {Alves}, {Anders}, {Anderson},
  {Anglada Varela}, {Antoja}, {Baines}, {Baker}, {Balaguer-N{\'u}{\~n}ez},
  {Balbinot}, {Balog}, {Barache}, {Barbato}, {Barros}, {Barstow},
  {Bartolom{\'e}}, {Bassilana}, {Bauchet}, {Becciani}, {Bellazzini},
  {Berihuete}, {Bernet}, {Bertone}, {Bianchi}, {Binnenfeld}, {Blanco-Cuaresma},
  {Boch}, {Bombrun}, {Bossini}, {Bouquillon}, {Bragaglia}, {Bramante},
  {Breedt}, {Bressan}, {Brouillet}, {Brugaletta}, {Bucciarelli}, {Burlacu},
  {Butkevich}, {Buzzi}, {Caffau}, {Cancelliere}, {Cantat-Gaudin}, {Carballo},
  {Carlucci}, {Carnerero}, {Carrasco}, {Casamiquela}, {Castellani},
  {Castro-Ginard}, {Chaoul}, {Charlot}, {Chemin}, {Chiaramida}, {Chiavassa},
  {Chornay}, {Comoretto}, {Contursi}, {Cooper}, {Cornez}, {Cowell}, {Crifo},
  {Cropper}, {Crosta}, {Crowley}, {Dafonte}, {Dapergolas}, {David}, {de
  Laverny}, {De Luise}, {De March}, {de Souza}, {de Torres}, {del Peloso}, {del
  Pozo}, {Delbo}, {Delgado}, {Delisle}, {Demouchy}, {Dharmawardena}, {Diakite},
  {Diener}, {Distefano}, {Dolding}, {Enke}, {Fabre}, {Fabrizio}, {Faigler},
  {Fedorets}, {Fernique}, {Figueras}, {Fournier}, {Fouron}, {Fragkoudi}, {Gai},
  {Garcia-Gutierrez}, {Garcia-Reinaldos}, {Garc{\'\i}a-Torres}, {Garofalo},
  {Gavel}, {Gavras}, {Gerlach}, {Geyer}, {Giacobbe}, {Gilmore}, {Girona},
  {Giuffrida}, {Gomel}, {Gomez}, {Gonz{\'a}lez-N{\'u}{\~n}ez},
  {Gonz{\'a}lez-Santamar{\'\i}a}, {Gonz{\'a}lez-Vidal}, {Granvik}, {Guillout},
  {Guiraud}, {Guti{\'e}rrez-S{\'a}nchez}, {Guy}, {Hatzidimitriou}, {Hauser},
  {Haywood}, {Helmer}, {Helmi}, {Sarmiento}, {Hidalgo}, {H{\l}adczuk}, {Hobbs},
  {Holland}, {Huckle}, {Jardine}, {Jasniewicz}, {Jean-Antoine Piccolo},
  {Jim{\'e}nez-Arranz}, {Juaristi Campillo}, {Julbe}, {Karbevska}, {Kervella},
  {Khanna}, {Kordopatis}, {Korn}, {K{\'o}sp{\'a}l}, {Kostrzewa-Rutkowska},
  {Kruszy{\'n}ska}, {Kun}, {Laizeau}, {Lambert}, {Lanza}, {Lasne}, {Le
  Campion}, {Lebreton}, {Lebzelter}, {Leccia}, {Leclerc}, {Lecoeur-Taibi},
  {Liao}, {Licata}, {Lindstr{\o}m}, {Lister}, {Livanou}, {Lobel}, {Lorca},
  {Loup}, {Madrero Pardo}, {Magdaleno Romeo}, {Managau}, {Mann}, {Manteiga},
  {Marchant}, {Marconi}, {Marcos}, {Marcos Santos}, {Mar{\'\i}n Pina},
  {Marinoni}, {Marocco}, {Marshall}, {Polo}, {Mart{\'\i}n-Fleitas}, {Marton},
  {Mary}, {Masip}, {Massari}, {Mastrobuono-Battisti}, {Mazeh}, {McMillan},
  {Messina}, {Michalik}, {Millar}, {Mints}, {Molina}, {Molinaro}, {Moln{\'a}r},
  {Monari}, {Mongui{\'o}}, {Montegriffo}, {Montero}, {Mor}, {Mora},
  {Morbidelli}, {Morel}, {Morris}, {Muraveva}, {Murphy}, {Musella}, {Nagy},
  {Noval}, {Oca{\~n}a}, {Ogden}, {Ordenovic}, {Osinde}, {Pagani}, {Pagano},
  {Palicio}, {Pallas-Quintela}, {Panahi}, {Payne-Wardenaar}, {Pe{\~n}alosa
  Esteller}, {Penttil{\"a}}, {Pichon}, {Piersimoni}, {Pineau}, {Plachy},
  {Plum}, {Poggio}, {Pr{\v{s}}a}, {Pulone}, {Racero}, {Ragaini}, {Rainer},
  {Raiteri}, {Ramos}, {Ramos-Lerate}, {Re Fiorentin}, {Regibo}, {Richards},
  {Rios Diaz}, {Riva}, {Rix}, {Rixon}, {Robichon}, {Robin}, {Robin}, {Roelens},
  {Rogues}, {Rohrbasser}, {Romero-G{\'o}mez}, {Rowell}, {Royer}, {Ruz Mieres},
  {Rybicki}, {Sadowski}, {S{\'a}ez N{\'u}{\~n}ez}, {Sagrist{\`a} Sell{\'e}s},
  {Sahlmann}, {Salguero}, {Samaras}, {Sanchez Gimenez}, {Sanna},
  {Santove{\~n}a}, {Sarasso}, {Schultheis}, {Sciacca}, {Segol}, {Segovia},
  {S{\'e}gransan}, {Semeux}, {Shahaf}, {Siddiqui}, {Siebert}, {Siltala},
  {Silvelo}, {Slezak}, {Slezak}, {Smart}, {Snaith}, {Solano}, {Solitro},
  {Souami}, {Souchay}, {Spagna}, {Spina}, {Spoto}, {Steele},
  {Steidelm{\"u}ller}, {Stephenson}, {S{\"u}veges}, {Surdej}, {Szabados},
  {Szegedi-Elek}, {Taris}, {Taylor}, {Teixeira}, {Tolomei}, {Tonello}, {Torra},
  {Torra}, {Torralba Elipe}, {Trabucchi}, {Tsounis}, {Turon}, {Ulla}, {Unger},
  {Vaillant}, {van Dillen}, {van Reeven}, {Vanel}, {Vecchiato}, {Viala},
  {Vicente}, {Voutsinas}, {Weiler}, {Wevers}, {Wyrzykowski}, {Yoldas}, {Yvard},
  {Zhao}, {Zorec}, {Zucker}, \& {Zwitter}}]{GaiaMSpuls2022}
{Gaia Collaboration}, {De Ridder}, J., {Ripepi}, V., {et~al.} 2022, arXiv
  e-prints, arXiv:2206.06075

\bibitem[{{Gaia Collaboration} {et~al.}(2019){Gaia Collaboration}, {Eyer},
  {Rimoldini}, {Audard}, {Anderson}, {Nienartowicz}, {Glass}, {Marchal},
  {Grenon}, {Mowlavi}, {Holl}, {Clementini}, {Aerts}, {Mazeh}, {Evans},
  {Szabados}, {Brown}, {Vallenari}, {Prusti}, {de Bruijne}, {Babusiaux},
  {Bailer-Jones}, {Biermann}, {Jansen}, {Jordi}, {Klioner}, {Lammers},
  {Lindegren}, {Luri}, {Mignard}, {Panem}, {Pourbaix}, {Randich}, {Sartoretti},
  {Siddiqui}, {Soubiran}, {van Leeuwen}, {Walton}, {Arenou}, {Bastian},
  {Cropper}, {Drimmel}, {Katz}, {Lattanzi}, {Bakker}, {Cacciari},
  {Casta{\~n}eda}, {Chaoul}, {Cheek}, {De Angeli}, {Fabricius}, {Guerra},
  {Masana}, {Messineo}, {Panuzzo}, {Portell}, {Riello}, {Seabroke}, {Tanga},
  {Th{\'e}venin}, {Gracia-Abril}, {Comoretto}, {Garcia-Reinaldos}, {Teyssier},
  {Altmann}, {Andrae}, {Bellas-Velidis}, {Benson}, {Berthier}, {Blomme},
  {Burgess}, {Busso}, {Carry}, {Cellino}, {Clotet}, {Creevey}, {Davidson}, {De
  Ridder}, {Delchambre}, {Dell'Oro}, {Ducourant},
  {Fern{\'a}ndez-Hern{\'a}ndez}, {Fouesneau}, {Fr{\'e}mat}, {Galluccio},
  {Garc{\'\i}a-Torres}, {Gonz{\'a}lez-N{\'u}{\~n}ez}, {Gonz{\'a}lez-Vidal},
  {Gosset}, {Guy}, {Halbwachs}, {Hambly}, {Harrison}, {Hern{\'a}ndez},
  {Hestroffer}, {Hodgkin}, {Hutton}, {Jasniewicz}, {Jean-Antoine-Piccolo},
  {Jordan}, {Korn}, {Krone-Martins}, {Lanzafame}, {Lebzelter}, {L{\"o}ffler},
  {Manteiga}, {Marrese}, {Mart{\'\i}n-Fleitas}, {Moitinho}, {Mora}, {Muinonen},
  {Osinde}, {Pancino}, {Pauwels}, {Petit}, {Recio-Blanco}, {Richards}, {Robin},
  {Sarro}, {Siopis}, {Smith}, {Sozzetti}, {S{\"u}veges}, {Torra}, {van Reeven},
  {Abbas}, {Abreu Aramburu}, {Accart}, {Altavilla}, {{\'A}lvarez}, {Alvarez},
  {Alves}, {Andrei}, {Anglada Varela}, {Antiche}, {Antoja}, {Arcay},
  {Astraatmadja}, {Bach}, {Baker}, {Balaguer-N{\'u}{\~n}ez}, {Balm}, {Barache},
  {Barata}, {Barbato}, {Barblan}, {Barklem}, {Barrado}, {Barros}, {Barstow},
  {Bartholom{\'e} Mu{\~n}oz}, {Bassilana}, {Becciani}, {Bellazzini},
  {Berihuete}, {Bertone}, {Bianchi}, {Bienaym{\'e}}, {Blanco-Cuaresma}, {Boch},
  {Boeche}, {Bombrun}, {Borrachero}, {Bossini}, {Bouquillon}, {Bourda},
  {Bragaglia}, {Bramante}, {Breddels}, {Bressan}, {Brouillet},
  {Br{\"u}semeister}, {Brugaletta}, {Bucciarelli}, {Burlacu}, {Busonero},
  {Butkevich}, {Buzzi}, {Caffau}, {Cancelliere}, {Cannizzaro}, {Cantat-Gaudin},
  {Carballo}, {Carlucci}, {Carrasco}, {Casamiquela}, {Castellani},
  {Castro-Ginard}, {Charlot}, {Chemin}, {Chiavassa}, {Cocozza}, {Costigan},
  {Cowell}, {Crifo}, {Crosta}, {Crowley}, {Cuypers}, {Dafonte}, {Damerdji},
  {Dapergolas}, {David}, {David}, {de Laverny}, {De Luise}, {De March}, {de
  Martino}, {de Souza}, {de Torres}, {Debosscher}, {del Pozo}, {Delbo},
  {Delgado}, {Delgado}, {Diakite}, {Diener}, {Distefano}, {Dolding},
  {Drazinos}, {Dur{\'a}n}, {Edvardsson}, {Enke}, {Eriksson}, {Esquej}, {Eynard
  Bontemps}, {Fabre}, {Fabrizio}, {Faigler}, {Falc{\~a}o}, {Farr{\`a}s Casas},
  {Federici}, {Fedorets}, {Fernique}, {Figueras}, {Filippi}, {Findeisen},
  {Fonti}, {Fraile}, {Fraser}, {Fr{\'e}zouls}, {Gai}, {Galleti}, {Garabato},
  {Garc{\'\i}a-Sedano}, {Garofalo}, {Garralda}, {Gavel}, {Gavras}, {Gerssen},
  {Geyer}, {Giacobbe}, {Gilmore}, {Girona}, {Giuffrida}, {Gomes}, {Granvik},
  {Gueguen}, {Guerrier}, {Guiraud}, {Guti{\'e}rrez-S{\'a}nchez}, {Haigron},
  {Hatzidimitriou}, {Hauser}, {Haywood}, {Heiter}, {Helmi}, {Heu}, {Hilger},
  {Hobbs}, {Hofmann}, {Holland}, {Huckle}, {Hypki}, {Icardi}, {Jan{\ss}en},
  {Jevardat de Fombelle}, {Jonker}, {Juh{\'a}sz}, {Julbe}, {Karampelas},
  {Kewley}, {Klar}, {Kochoska}, {Kohley}, {Kolenberg}, {Kontizas}, {Kontizas},
  {Koposov}, {Kordopatis}, {Kostrzewa-Rutkowska}, {Koubsky}, {Lambert},
  {Lanza}, {Lasne}, {Lavigne}, {Le Fustec}, {Le Poncin-Lafitte}, {Lebreton},
  {Leccia}, {Leclerc}, {Lecoeur-Taibi}, {Lenhardt}, {Leroux}, {Liao}, {Licata},
  {Lindstr{\o}m}, {Lister}, {Livanou}, {Lobel}, {L{\'o}pez}, {Lorenz},
  {Managau}, {Mann}, {Mantelet}, {Marchant}, {Marconi}, {Marinoni},
  {Marschalk{\'o}}, {Marshall}, {Martino}, {Marton}, {Mary}, {Massari},
  {Matijevi{\v{c}}}, {McMillan}, {Messina}, {Michalik}, {Millar}, {Molina},
  {Molinaro}, {Moln{\'a}r}, {Montegriffo}, {Mor}, {Morbidelli}, {Morel},
  {Morgenthaler}, {Morris}, {Mulone}, {Muraveva}, {Musella}, {Nelemans},
  {Nicastro}, {Noval}, {O'Mullane}, {Ord{\'e}novic}, {Ord{\'o}{\~n}ez-Blanco},
  {Osborne}, {Pagani}, {Pagano}, {Pailler}, {Palacin}, {Palaversa}, {Panahi},
  {Pawlak}, {Piersimoni}, {Pineau}, {Plachy}, {Plum}, {Poggio}, {Poujoulet},
  {Pr{\v{s}}a}, {Pulone}, {Racero}, {Ragaini}, {Rambaux}, {Ramos-Lerate},
  {Regibo}, {Reyl{\'e}}, {Riclet}, {Ripepi}, {Riva}, {Rivard}, {Rixon},
  {Roegiers}, {Roelens}, {Romero-G{\'o}mez}, {Rowell}, {Royer}, {Ruiz-Dern},
  {Sadowski}, {Sagrist{\`a} Sell{\'e}s}, {Sahlmann}, {Salgado}, {Salguero},
  {Sanna}, {Santana-Ros}, {Sarasso}, {Savietto}, {Schultheis}, {Sciacca},
  {Segol}, {Segovia}, {S{\'e}gransan}, {Shih}, {Siltala}, {Silva}, {Smart},
  {Smith}, {Solano}, {Solitro}, {Sordo}, {Soria Nieto}, {Souchay}, {Spagna},
  {Spoto}, {Stampa}, {Steele}, {Steidelm{\"u}ller}, {Stephenson}, {Stoev},
  {Suess}, {Surdej}, {Szegedi-Elek}, {Tapiador}, {Taris}, {Tauran}, {Taylor},
  {Teixeira}, {Terrett}, {Teyssandier}, {Thuillot}, {Titarenko}, {Torra
  Clotet}, {Turon}, {Ulla}, {Utrilla}, {Uzzi}, {Vaillant}, {Valentini},
  {Valette}, {van Elteren}, {Van Hemelryck}, {van Leeuwen}, {Vaschetto},
  {Vecchiato}, {Veljanoski}, {Viala}, {Vicente}, {Vogt}, {von Essen}, {Voss},
  {Votruba}, {Voutsinas}, {Walmsley}, {Weiler}, {Wertz}, {Wevers},
  {Wyrzykowski}, {Yoldas}, {{\v{Z}}erjal}, {Ziaeepour}, {Zorec}, {Zschocke},
  {Zucker}, {Zurbach}, \& {Zwitter}}]{DR2puls}
{Gaia Collaboration}, {Eyer}, L., {Rimoldini}, L., {et~al.} 2019, \aap, 623,
  A110

\bibitem[{{Gaia Collaboration} {et~al.}(2021{\natexlab{b}}){Gaia
  Collaboration}, {Luri}, {Chemin}, {Clementini}, {Delgado}, {McMillan},
  {Romero-G{\'o}mez}, {Balbinot}, {Castro-Ginard}, {Mor}, {Ripepi}, {Sarro},
  {Cioni}, {Fabricius}, {Garofalo}, {Helmi}, {Muraveva}, {Brown}, {Vallenari},
  {Prusti}, {de Bruijne}, {Babusiaux}, {Biermann}, {Creevey}, {Evans}, {Eyer},
  {Hutton}, {Jansen}, {Jordi}, {Klioner}, {Lammers}, {Lindegren}, {Mignard},
  {Panem}, {Pourbaix}, {Randich}, {Sartoretti}, {Soubiran}, {Walton}, {Arenou},
  {Bailer-Jones}, {Bastian}, {Cropper}, {Drimmel}, {Katz}, {Lattanzi}, {van
  Leeuwen}, {Bakker}, {Casta{\~n}eda}, {De Angeli}, {Ducourant}, {Fouesneau},
  {Fr{\'e}mat}, {Guerra}, {Guerrier}, {Guiraud}, {Jean-Antoine Piccolo},
  {Masana}, {Messineo}, {Mowlavi}, {Nicolas}, {Nienartowicz}, {Pailler},
  {Panuzzo}, {Riclet}, {Roux}, {Seabroke}, {Sordo}, {Tanga}, {Th{\'e}venin},
  {Gracia-Abril}, {Portell}, {Teyssier}, {Altmann}, {Andrae}, {Bellas-Velidis},
  {Benson}, {Berthier}, {Blomme}, {Brugaletta}, {Burgess}, {Busso}, {Carry},
  {Cellino}, {Cheek}, {Damerdji}, {Davidson}, {Delchambre}, {Dell'Oro},
  {Fern{\'a}ndez-Hern{\'a}ndez}, {Galluccio}, {Garc{\'\i}a-Lario},
  {Garcia-Reinaldos}, {Gonz{\'a}lez-N{\'u}{\~n}ez}, {Gosset}, {Haigron},
  {Halbwachs}, {Hambly}, {Harrison}, {Hatzidimitriou}, {Heiter},
  {Hern{\'a}ndez}, {Hestroffer}, {Hodgkin}, {Holl}, {Jan{\ss}en}, {Jevardat de
  Fombelle}, {Jordan}, {Krone-Martins}, {Lanzafame}, {L{\"o}ffler}, {Lorca},
  {Manteiga}, {Marchal}, {Marrese}, {Moitinho}, {Mora}, {Muinonen}, {Osborne},
  {Pancino}, {Pauwels}, {Recio-Blanco}, {Richards}, {Riello}, {Rimoldini},
  {Robin}, {Roegiers}, {Rybizki}, {Siopis}, {Smith}, {Sozzetti}, {Ulla},
  {Utrilla}, {van Leeuwen}, {van Reeven}, {Abbas}, {Abreu Aramburu}, {Accart},
  {Aerts}, {Aguado}, {Ajaj}, {Altavilla}, {{\'A}lvarez}, {{\'A}lvarez
  Cid-Fuentes}, {Alves}, {Anderson}, {Anglada Varela}, {Antoja}, {Audard},
  {Baines}, {Baker}, {Balaguer-N{\'u}{\~n}ez}, {Balog}, {Barache}, {Barbato},
  {Barros}, {Barstow}, {Bartolom{\'e}}, {Bassilana}, {Bauchet},
  {Baudesson-Stella}, {Becciani}, {Bellazzini}, {Bernet}, {Bertone}, {Bianchi},
  {Blanco-Cuaresma}, {Boch}, {Bombrun}, {Bossini}, {Bouquillon}, {Bragaglia},
  {Bramante}, {Breedt}, {Bressan}, {Brouillet}, {Bucciarelli}, {Burlacu},
  {Busonero}, {Butkevich}, {Buzzi}, {Caffau}, {Cancelliere}, {C{\'a}novas},
  {Cantat-Gaudin}, {Carballo}, {Carlucci}, {Carnerero}, {Carrasco},
  {Casamiquela}, {Castellani}, {Castro Sampol}, {Chaoul}, {Charlot},
  {Chiavassa}, {Comoretto}, {Cooper}, {Cornez}, {Cowell}, {Crifo}, {Crosta},
  {Crowley}, {Dafonte}, {Dapergolas}, {David}, {David}, {de Laverny}, {De
  Luise}, {De March}, {De Ridder}, {de Souza}, {de Teodoro}, {de Torres}, {del
  Peloso}, {del Pozo}, {Delgado}, {Delisle}, {Di Matteo}, {Diakite}, {Diener},
  {Distefano}, {Dolding}, {Eappachen}, {Enke}, {Esquej}, {Fabre}, {Fabrizio},
  {Faigler}, {Fedorets}, {Fernique}, {Fienga}, {Figueras}, {Fouron},
  {Fragkoudi}, {Fraile}, {Franke}, {Gai}, {Garabato}, {Garcia-Gutierrez},
  {Garc{\'\i}a-Torres}, {Gavras}, {Gerlach}, {Geyer}, {Giacobbe}, {Gilmore},
  {Girona}, {Giuffrida}, {Gomez}, {Gonzalez-Santamaria}, {Gonz{\'a}lez-Vidal},
  {Granvik}, {Guti{\'e}rrez-S{\'a}nchez}, {Guy}, {Hauser}, {Haywood},
  {Hidalgo}, {Hilger}, {H{\l}adczuk}, {Hobbs}, {Holland}, {Huckle},
  {Jasniewicz}, {Jonker}, {Juaristi Campillo}, {Julbe}, {Karbevska},
  {Kervella}, {Khanna}, {Kochoska}, {Kontizas}, {Kordopatis}, {Korn},
  {Kostrzewa-Rutkowska}, {Kruszy{\'n}ska}, {Lambert}, {Lanza}, {Lasne}, {Le
  Campion}, {Le Fustec}, {Lebreton}, {Lebzelter}, {Leccia}, {Leclerc},
  {Lecoeur-Taibi}, {Liao}, {Licata}, {Lindstr{\o}m}, {Lister}, {Livanou},
  {Lobel}, {Madrero Pardo}, {Managau}, {Mann}, {Marchant}, {Marconi}, {Marcos
  Santos}, {Marinoni}, {Marocco}, {Marshall}, {Martin Polo},
  {Mart{\'\i}n-Fleitas}, {Masip}, {Massari}, {Mastrobuono-Battisti}, {Mazeh},
  {Messina}, {Michalik}, {Millar}, {Mints}, {Molina}, {Molinaro}, {Moln{\'a}r},
  {Montegriffo}, {Morbidelli}, {Morel}, {Morris}, {Mulone}, {Munoz}, {Murphy},
  {Musella}, {Noval}, {Ord{\'e}novic}, {Orr{\`u}}, {Osinde}, {Pagani},
  {Pagano}, {Palaversa}, {Palicio}, {Panahi}, {Pawlak}, {Pe{\~n}alosa
  Esteller}, {Penttil{\"a}}, {Piersimoni}, {Pineau}, {Plachy}, {Plum},
  {Poggio}, {Poretti}, {Poujoulet}, {Pr{\v{s}}a}, {Pulone}, {Racero},
  {Ragaini}, {Rainer}, {Raiteri}, {Rambaux}, {Ramos}, {Ramos-Lerate}, {Re
  Fiorentin}, {Regibo}, {Reyl{\'e}}, {Riva}, {Rixon}, {Robichon}, {Robin},
  {Roelens}, {Rohrbasser}, {Rowell}, {Royer}, {Rybicki}, {Sadowski},
  {Sagrist{\`a} Sell{\'e}s}, {Sahlmann}, {Salgado}, {Salguero}, {Samaras},
  {Gimenez}, {Sanna}, {Santove{\~n}a}, {Sarasso}, {Schultheis}, {Sciacca},
  {Segol}, {Segovia}, {S{\'e}gransan}, {Semeux}, {Siddiqui}, {Siebert},
  {Siltala}, {Slezak}, {Smart}, {Solano}, {Solitro}, {Souami}, {Souchay},
  {Spagna}, {Spoto}, {Steele}, {Steidelm{\"u}ller}, {Stephenson},
  {S{\"u}veges}, {Szabados}, {Szegedi-Elek}, {Taris}, {Tauran}, {Taylor},
  {Teixeira}, {Thuillot}, {Tonello}, {Torra}, {Torra}, {Turon}, {Unger},
  {Vaillant}, {van Dillen}, {Vanel}, {Vecchiato}, {Viala}, {Vicente},
  {Voutsinas}, {Weiler}, {Wevers}, {Wyrzykowski}, {Yoldas}, {Yvard}, {Zhao},
  {Zorec}, {Zucker}, {Zurbach}, \& {Zwitter}}]{Luri2021}
{Gaia Collaboration}, {Luri}, X., {Chemin}, L., {et~al.} 2021{\natexlab{b}},
  \aap, 649, A7

\bibitem[{{Gaia Collaboration} {et~al.}(2016){Gaia Collaboration}, {Prusti},
  {de Bruijne}, {Brown}, {Vallenari}, {Babusiaux}, {Bailer-Jones}, {Bastian},
  {Biermann}, {Evans}, {Eyer}, {Jansen}, {Jordi}, {Klioner}, {Lammers},
  {Lindegren}, {Luri}, {Mignard}, {Milligan}, {Panem}, {Poinsignon},
  {Pourbaix}, {Randich}, {Sarri}, {Sartoretti}, {Siddiqui}, {Soubiran},
  {Valette}, {van Leeuwen}, {Walton}, {Aerts}, {Arenou}, {Cropper}, {Drimmel},
  {H{\o}g}, {Katz}, {Lattanzi}, {O'Mullane}, {Grebel}, {Holland}, {Huc},
  {Passot}, {Bramante}, {Cacciari}, {Casta{\~n}eda}, {Chaoul}, {Cheek}, {De
  Angeli}, {Fabricius}, {Guerra}, {Hern{\'a}ndez}, {Jean-Antoine-Piccolo},
  {Masana}, {Messineo}, {Mowlavi}, {Nienartowicz}, {Ord{\'o}{\~n}ez-Blanco},
  {Panuzzo}, {Portell}, {Richards}, {Riello}, {Seabroke}, {Tanga},
  {Th{\'e}venin}, {Torra}, {Els}, {Gracia-Abril}, {Comoretto},
  {Garcia-Reinaldos}, {Lock}, {Mercier}, {Altmann}, {Andrae}, {Astraatmadja},
  {Bellas-Velidis}, {Benson}, {Berthier}, {Blomme}, {Busso}, {Carry},
  {Cellino}, {Clementini}, {Cowell}, {Creevey}, {Cuypers}, {Davidson}, {De
  Ridder}, {de Torres}, {Delchambre}, {Dell'Oro}, {Ducourant}, {Fr{\'e}mat},
  {Garc{\'\i}a-Torres}, {Gosset}, {Halbwachs}, {Hambly}, {Harrison}, {Hauser},
  {Hestroffer}, {Hodgkin}, {Huckle}, {Hutton}, {Jasniewicz}, {Jordan},
  {Kontizas}, {Korn}, {Lanzafame}, {Manteiga}, {Moitinho}, {Muinonen},
  {Osinde}, {Pancino}, {Pauwels}, {Petit}, {Recio-Blanco}, {Robin}, {Sarro},
  {Siopis}, {Smith}, {Smith}, {Sozzetti}, {Thuillot}, {van Reeven}, {Viala},
  {Abbas}, {Abreu Aramburu}, {Accart}, {Aguado}, {Allan}, {Allasia},
  {Altavilla}, {{\'A}lvarez}, {Alves}, {Anderson}, {Andrei}, {Anglada Varela},
  {Antiche}, {Antoja}, {Ant{\'o}n}, {Arcay}, {Atzei}, {Ayache}, {Bach},
  {Baker}, {Balaguer-N{\'u}{\~n}ez}, {Barache}, {Barata}, {Barbier}, {Barblan},
  {Baroni}, {Barrado y Navascu{\'e}s}, {Barros}, {Barstow}, {Becciani},
  {Bellazzini}, {Bellei}, {Bello Garc{\'\i}a}, {Belokurov}, {Bendjoya},
  {Berihuete}, {Bianchi}, {Bienaym{\'e}}, {Billebaud}, {Blagorodnova},
  {Blanco-Cuaresma}, {Boch}, {Bombrun}, {Borrachero}, {Bouquillon}, {Bourda},
  {Bouy}, {Bragaglia}, {Breddels}, {Brouillet}, {Br{\"u}semeister},
  {Bucciarelli}, {Budnik}, {Burgess}, {Burgon}, {Burlacu}, {Busonero}, {Buzzi},
  {Caffau}, {Cambras}, {Campbell}, {Cancelliere}, {Cantat-Gaudin}, {Carlucci},
  {Carrasco}, {Castellani}, {Charlot}, {Charnas}, {Charvet}, {Chassat},
  {Chiavassa}, {Clotet}, {Cocozza}, {Collins}, {Collins}, {Costigan}, {Crifo},
  {Cross}, {Crosta}, {Crowley}, {Dafonte}, {Damerdji}, {Dapergolas}, {David},
  {David}, {De Cat}, {de Felice}, {de Laverny}, {De Luise}, {De March}, {de
  Martino}, {de Souza}, {Debosscher}, {del Pozo}, {Delbo}, {Delgado},
  {Delgado}, {di Marco}, {Di Matteo}, {Diakite}, {Distefano}, {Dolding}, {Dos
  Anjos}, {Drazinos}, {Dur{\'a}n}, {Dzigan}, {Ecale}, {Edvardsson}, {Enke},
  {Erdmann}, {Escolar}, {Espina}, {Evans}, {Eynard Bontemps}, {Fabre},
  {Fabrizio}, {Faigler}, {Falc{\~a}o}, {Farr{\`a}s Casas}, {Faye}, {Federici},
  {Fedorets}, {Fern{\'a}ndez-Hern{\'a}ndez}, {Fernique}, {Fienga}, {Figueras},
  {Filippi}, {Findeisen}, {Fonti}, {Fouesneau}, {Fraile}, {Fraser}, {Fuchs},
  {Furnell}, {Gai}, {Galleti}, {Galluccio}, {Garabato}, {Garc{\'\i}a-Sedano},
  {Gar{\'e}}, {Garofalo}, {Garralda}, {Gavras}, {Gerssen}, {Geyer}, {Gilmore},
  {Girona}, {Giuffrida}, {Gomes}, {Gonz{\'a}lez-Marcos},
  {Gonz{\'a}lez-N{\'u}{\~n}ez}, {Gonz{\'a}lez-Vidal}, {Granvik}, {Guerrier},
  {Guillout}, {Guiraud}, {G{\'u}rpide}, {Guti{\'e}rrez-S{\'a}nchez}, {Guy},
  {Haigron}, {Hatzidimitriou}, {Haywood}, {Heiter}, {Helmi}, {Hobbs},
  {Hofmann}, {Holl}, {Holland}, {Hunt}, {Hypki}, {Icardi}, {Irwin}, {Jevardat
  de Fombelle}, {Jofr{\'e}}, {Jonker}, {Jorissen}, {Julbe}, {Karampelas},
  {Kochoska}, {Kohley}, {Kolenberg}, {Kontizas}, {Koposov}, {Kordopatis},
  {Koubsky}, {Kowalczyk}, {Krone-Martins}, {Kudryashova}, {Kull}, {Bachchan},
  {Lacoste-Seris}, {Lanza}, {Lavigne}, {Le Poncin-Lafitte}, {Lebreton},
  {Lebzelter}, {Leccia}, {Leclerc}, {Lecoeur-Taibi}, {Lemaitre}, {Lenhardt},
  {Leroux}, {Liao}, {Licata}, {Lindstr{\o}m}, {Lister}, {Livanou}, {Lobel},
  {L{\"o}ffler}, {L{\'o}pez}, {Lopez-Lozano}, {Lorenz}, {Loureiro},
  {MacDonald}, {Magalh{\~a}es Fernandes}, {Managau}, {Mann}, {Mantelet},
  {Marchal}, {Marchant}, {Marconi}, {Marie}, {Marinoni}, {Marrese},
  {Marschalk{\'o}}, {Marshall}, {Mart{\'\i}n-Fleitas}, {Martino}, {Mary},
  {Matijevi{\v{c}}}, {Mazeh}, {McMillan}, {Messina}, {Mestre}, {Michalik},
  {Millar}, {Miranda}, {Molina}, {Molinaro}, {Molinaro}, {Moln{\'a}r},
  {Moniez}, {Montegriffo}, {Monteiro}, {Mor}, {Mora}, {Morbidelli}, {Morel},
  {Morgenthaler}, {Morley}, {Morris}, {Mulone}, {Muraveva}, {Musella},
  {Narbonne}, {Nelemans}, {Nicastro}, {Noval}, {Ord{\'e}novic},
  {Ordieres-Mer{\'e}}, {Osborne}, {Pagani}, {Pagano}, {Pailler}, {Palacin},
  {Palaversa}, {Parsons}, {Paulsen}, {Pecoraro}, {Pedrosa}, {Pentik{\"a}inen},
  {Pereira}, {Pichon}, {Piersimoni}, {Pineau}, {Plachy}, {Plum}, {Poujoulet},
  {Pr{\v{s}}a}, {Pulone}, {Ragaini}, {Rago}, {Rambaux}, {Ramos-Lerate},
  {Ranalli}, {Rauw}, {Read}, {Regibo}, {Renk}, {Reyl{\'e}}, {Ribeiro},
  {Rimoldini}, {Ripepi}, {Riva}, {Rixon}, {Roelens}, {Romero-G{\'o}mez},
  {Rowell}, {Royer}, {Rudolph}, {Ruiz-Dern}, {Sadowski}, {Sagrist{\`a}
  Sell{\'e}s}, {Sahlmann}, {Salgado}, {Salguero}, {Sarasso}, {Savietto},
  {Schnorhk}, {Schultheis}, {Sciacca}, {Segol}, {Segovia}, {Segransan},
  {Serpell}, {Shih}, {Smareglia}, {Smart}, {Smith}, {Solano}, {Solitro},
  {Sordo}, {Soria Nieto}, {Souchay}, {Spagna}, {Spoto}, {Stampa}, {Steele},
  {Steidelm{\"u}ller}, {Stephenson}, {Stoev}, {Suess}, {S{\"u}veges}, {Surdej},
  {Szabados}, {Szegedi-Elek}, {Tapiador}, {Taris}, {Tauran}, {Taylor},
  {Teixeira}, {Terrett}, {Tingley}, {Trager}, {Turon}, {Ulla}, {Utrilla},
  {Valentini}, {van Elteren}, {Van Hemelryck}, {van Leeuwen}, {Varadi},
  {Vecchiato}, {Veljanoski}, {Via}, {Vicente}, {Vogt}, {Voss}, {Votruba},
  {Voutsinas}, {Walmsley}, {Weiler}, {Weingrill}, {Werner}, {Wevers},
  {Whitehead}, {Wyrzykowski}, {Yoldas}, {{\v{Z}}erjal}, {Zucker}, {Zurbach},
  {Zwitter}, {Alecu}, {Allen}, {Allende Prieto}, {Amorim},
  {Anglada-Escud{\'e}}, {Arsenijevic}, {Azaz}, {Balm}, {Beck}, {Bernstein},
  {Bigot}, {Bijaoui}, {Blasco}, {Bonfigli}, {Bono}, {Boudreault}, {Bressan},
  {Brown}, {Brunet}, {Bunclark}, {Buonanno}, {Butkevich}, {Carret}, {Carrion},
  {Chemin}, {Ch{\'e}reau}, {Corcione}, {Darmigny}, {de Boer}, {de Teodoro}, {de
  Zeeuw}, {Delle Luche}, {Domingues}, {Dubath}, {Fodor}, {Fr{\'e}zouls},
  {Fries}, {Fustes}, {Fyfe}, {Gallardo}, {Gallegos}, {Gardiol}, {Gebran},
  {Gomboc}, {G{\'o}mez}, {Grux}, {Gueguen}, {Heyrovsky}, {Hoar}, {Iannicola},
  {Isasi Parache}, {Janotto}, {Joliet}, {Jonckheere}, {Keil}, {Kim},
  {Klagyivik}, {Klar}, {Knude}, {Kochukhov}, {Kolka}, {Kos}, {Kutka}, {Lainey},
  {LeBouquin}, {Liu}, {Loreggia}, {Makarov}, {Marseille}, {Martayan},
  {Martinez-Rubi}, {Massart}, {Meynadier}, {Mignot}, {Munari}, {Nguyen},
  {Nordlander}, {Ocvirk}, {O'Flaherty}, {Olias Sanz}, {Ortiz}, {Osorio},
  {Oszkiewicz}, {Ouzounis}, {Palmer}, {Park}, {Pasquato}, {Peltzer}, {Peralta},
  {P{\'e}turaud}, {Pieniluoma}, {Pigozzi}, {Poels}, {Prat}, {Prod'homme},
  {Raison}, {Rebordao}, {Risquez}, {Rocca-Volmerange}, {Rosen}, {Ruiz-Fuertes},
  {Russo}, {Sembay}, {Serraller Vizcaino}, {Short}, {Siebert}, {Silva},
  {Sinachopoulos}, {Slezak}, {Soffel}, {Sosnowska}, {Strai{\v{z}}ys}, {ter
  Linden}, {Terrell}, {Theil}, {Tiede}, {Troisi}, {Tsalmantza}, {Tur},
  {Vaccari}, {Vachier}, {Valles}, {Van Hamme}, {Veltz}, {Virtanen}, {Wallut},
  {Wichmann}, {Wilkinson}, {Ziaeepour}, \& {Zschocke}}]{Gaiamission2016}
{Gaia Collaboration}, {Prusti}, T., {de Bruijne}, J.~H.~J., {et~al.} 2016,
  \aap, 595, A1

\bibitem[{{Gaia Collaboration} {et~al.}(2021{\natexlab{c}}){Gaia
  Collaboration}, {Smart}, {Sarro}, {Rybizki}, {Reyl{\'e}}, {Robin}, {Hambly},
  {Abbas}, {Barstow}, {de Bruijne}, {Bucciarelli}, {Carrasco}, {Cooper},
  {Hodgkin}, {Masana}, {Michalik}, {Sahlmann}, {Sozzetti}, {Brown},
  {Vallenari}, {Prusti}, {Babusiaux}, {Biermann}, {Creevey}, {Evans}, {Eyer},
  {Hutton}, {Jansen}, {Jordi}, {Klioner}, {Lammers}, {Lindegren}, {Luri},
  {Mignard}, {Panem}, {Pourbaix}, {Randich}, {Sartoretti}, {Soubiran},
  {Walton}, {Arenou}, {Bailer-Jones}, {Bastian}, {Cropper}, {Drimmel}, {Katz},
  {Lattanzi}, {van Leeuwen}, {Bakker}, {Casta{\~n}eda}, {De Angeli},
  {Ducourant}, {Fabricius}, {Fouesneau}, {Fr{\'e}mat}, {Guerra}, {Guerrier},
  {Guiraud}, {Jean-Antoine Piccolo}, {Messineo}, {Mowlavi}, {Nicolas},
  {Nienartowicz}, {Pailler}, {Panuzzo}, {Riclet}, {Roux}, {Seabroke}, {Sordo},
  {Tanga}, {Th{\'e}venin}, {Gracia-Abril}, {Portell}, {Teyssier}, {Altmann},
  {Andrae}, {Bellas-Velidis}, {Benson}, {Berthier}, {Blomme}, {Brugaletta},
  {Burgess}, {Busso}, {Carry}, {Cellino}, {Cheek}, {Clementini}, {Damerdji},
  {Davidson}, {Delchambre}, {Dell'Oro}, {Fern{\'a}ndez-Hern{\'a}ndez},
  {Galluccio}, {Garc{\'\i}a-Lario}, {Garcia-Reinaldos},
  {Gonz{\'a}lez-N{\'u}{\~n}ez}, {Gosset}, {Haigron}, {Halbwachs}, {Harrison},
  {Hatzidimitriou}, {Heiter}, {Hern{\'a}ndez}, {Hestroffer}, {Holl},
  {Jan{\ss}en}, {Jevardat de Fombelle}, {Jordan}, {Krone-Martins}, {Lanzafame},
  {L{\"o}ffler}, {Lorca}, {Manteiga}, {Marchal}, {Marrese}, {Moitinho}, {Mora},
  {Muinonen}, {Osborne}, {Pancino}, {Pauwels}, {Recio-Blanco}, {Richards},
  {Riello}, {Rimoldini}, {Roegiers}, {Siopis}, {Smith}, {Ulla}, {Utrilla}, {van
  Leeuwen}, {van Reeven}, {Abreu Aramburu}, {Accart}, {Aerts}, {Aguado},
  {Ajaj}, {Altavilla}, {{\'A}lvarez}, {{\'A}lvarez Cid-Fuentes}, {Alves},
  {Anderson}, {Anglada Varela}, {Antoja}, {Audard}, {Baines}, {Baker},
  {Balaguer-N{\'u}{\~n}ez}, {Balbinot}, {Balog}, {Barache}, {Barbato},
  {Barros}, {Bartolom{\'e}}, {Bassilana}, {Bauchet}, {Baudesson-Stella},
  {Becciani}, {Bellazzini}, {Bernet}, {Bertone}, {Bianchi}, {Blanco-Cuaresma},
  {Boch}, {Bombrun}, {Bossini}, {Bouquillon}, {Bragaglia}, {Bramante},
  {Breedt}, {Bressan}, {Brouillet}, {Burlacu}, {Busonero}, {Butkevich},
  {Buzzi}, {Caffau}, {Cancelliere}, {C{\'a}novas}, {Cantat-Gaudin}, {Carballo},
  {Carlucci}, {Carnerero}, {Casamiquela}, {Castellani}, {Castro-Ginard},
  {Castro Sampol}, {Chaoul}, {Charlot}, {Chemin}, {Chiavassa}, {Cioni},
  {Comoretto}, {Cornez}, {Cowell}, {Crifo}, {Crosta}, {Crowley}, {Dafonte},
  {Dapergolas}, {David}, {David}, {de Laverny}, {De Luise}, {De March}, {De
  Ridder}, {de Souza}, {de Teodoro}, {de Torres}, {del Peloso}, {del Pozo},
  {Delgado}, {Delgado}, {Delisle}, {Di Matteo}, {Diakite}, {Diener},
  {Distefano}, {Dolding}, {Eappachen}, {Edvardsson}, {Enke}, {Esquej}, {Fabre},
  {Fabrizio}, {Faigler}, {Fedorets}, {Fernique}, {Fienga}, {Figueras},
  {Fouron}, {Fragkoudi}, {Fraile}, {Franke}, {Gai}, {Garabato},
  {Garcia-Gutierrez}, {Garc{\'\i}a-Torres}, {Garofalo}, {Gavras}, {Gerlach},
  {Geyer}, {Giacobbe}, {Gilmore}, {Girona}, {Giuffrida}, {Gomel}, {Gomez},
  {Gonzalez-Santamaria}, {Gonz{\'a}lez-Vidal}, {Granvik},
  {Guti{\'e}rrez-S{\'a}nchez}, {Guy}, {Hauser}, {Haywood}, {Helmi}, {Hidalgo},
  {Hilger}, {H{\l}adczuk}, {Hobbs}, {Holland}, {Huckle}, {Jasniewicz},
  {Jonker}, {Juaristi Campillo}, {Julbe}, {Karbevska}, {Kervella}, {Khanna},
  {Kochoska}, {Kontizas}, {Kordopatis}, {Korn}, {Kostrzewa-Rutkowska},
  {Kruszy{\'n}ska}, {Lambert}, {Lanza}, {Lasne}, {Le Campion}, {Le Fustec},
  {Lebreton}, {Lebzelter}, {Leccia}, {Leclerc}, {Lecoeur-Taibi}, {Liao},
  {Licata}, {Lindstr{\o}m}, {Lister}, {Livanou}, {Lobel}, {Madrero Pardo},
  {Managau}, {Mann}, {Marchant}, {Marconi}, {Marcos Santos}, {Marinoni},
  {Marocco}, {Marshall}, {Martin Polo}, {Mart{\'\i}n-Fleitas}, {Masip},
  {Massari}, {Mastrobuono-Battisti}, {Mazeh}, {McMillan}, {Messina}, {Millar},
  {Mints}, {Molina}, {Molinaro}, {Moln{\'a}r}, {Montegriffo}, {Mor},
  {Morbidelli}, {Morel}, {Morris}, {Mulone}, {Munoz}, {Muraveva}, {Murphy},
  {Musella}, {Noval}, {Ord{\'e}novic}, {Orr{\`u}}, {Osinde}, {Pagani},
  {Pagano}, {Palaversa}, {Palicio}, {Panahi}, {Pawlak}, {Pe{\~n}alosa
  Esteller}, {Penttil{\"a}}, {Piersimoni}, {Pineau}, {Plachy}, {Plum},
  {Poggio}, {Poretti}, {Poujoulet}, {Pr{\v{s}}a}, {Pulone}, {Racero},
  {Ragaini}, {Rainer}, {Raiteri}, {Rambaux}, {Ramos}, {Ramos-Lerate}, {Re
  Fiorentin}, {Regibo}, {Ripepi}, {Riva}, {Rixon}, {Robichon}, {Robin},
  {Roelens}, {Rohrbasser}, {Romero-G{\'o}mez}, {Rowell}, {Royer}, {Rybicki},
  {Sadowski}, {Sagrist{\`a} Sell{\'e}s}, {Salgado}, {Salguero}, {Samaras},
  {Sanchez Gimenez}, {Sanna}, {Santove{\~n}a}, {Sarasso}, {Schultheis},
  {Sciacca}, {Segol}, {Segovia}, {S{\'e}gransan}, {Semeux}, {Shahaf},
  {Siddiqui}, {Siebert}, {Siltala}, {Slezak}, {Solano}, {Solitro}, {Souami},
  {Souchay}, {Spagna}, {Spoto}, {Steele}, {Steidelm{\"u}ller}, {Stephenson},
  {S{\"u}veges}, {Szabados}, {Szegedi-Elek}, {Taris}, {Tauran}, {Taylor},
  {Teixeira}, {Thuillot}, {Tonello}, {Torra}, {Torra}, {Turon}, {Unger},
  {Vaillant}, {van Dillen}, {Vanel}, {Vecchiato}, {Viala}, {Vicente},
  {Voutsinas}, {Weiler}, {Wevers}, {Wyrzykowski}, {Yoldas}, {Yvard}, {Zhao},
  {Zorec}, {Zucker}, {Zurbach}, \& {Zwitter}}]{GCNS2021}
{Gaia Collaboration}, {Smart}, R.~L., {Sarro}, L.~M., {et~al.}
  2021{\natexlab{c}}, \aap, 649, A6

\bibitem[{{Ginsburg} {et~al.}(2019){Ginsburg}, {Sip{\H o}cz}, {Brasseur},
  {Cowperthwaite}, {Craig}, {Deil}, {Guillochon}, {Guzman}, {Liedtke}, {Lian
  Lim}, {Lockhart}, {Mommert}, {Morris}, {Norman}, {Parikh}, {Persson},
  {Robitaille}, {Segovia}, {Singer}, {Tollerud}, {de Val-Borro}, {Valtchanov},
  {Woillez}, {The Astroquery collaboration}, \& {a subset of the astropy
  collaboration}}]{astroquery}
{Ginsburg}, A., {Sip{\H o}cz}, B.~M., {Brasseur}, C.~E., {et~al.} 2019, \aj,
  157, 98

\bibitem[{{Green} {et~al.}(2019){Green}, {Schlafly}, {Zucker}, {Speagle}, \&
  {Finkbeiner}}]{Greenetal2019}
{Green}, G.~M., {Schlafly}, E., {Zucker}, C., {Speagle}, J.~S., \&
  {Finkbeiner}, D. 2019, \apj, 887, 93

\bibitem[{{Jeon} {et~al.}(2004){Jeon}, {Lee}, {Kim}, \& {Lee}}]{Jeonetal2004}
{Jeon}, Y.-B., {Lee}, M.~G., {Kim}, S.-L., \& {Lee}, H. 2004, \aj, 128, 287

\bibitem[{{Kaye} {et~al.}(1999){Kaye}, {Handler}, {Krisciunas}, {Poretti}, \&
  {Zerbi}}]{Kayeetal1999}
{Kaye}, A.~B., {Handler}, G., {Krisciunas}, K., {Poretti}, E., \& {Zerbi},
  F.~M. 1999, \pasp, 111, 840

\bibitem[{{Lenz} \& {Breger}(2005)}]{LenzandBreger2005}
{Lenz}, P. \& {Breger}, M. 2005, Communications in Asteroseismology, 146, 53

\bibitem[{{Moln{\'a}r} {et~al.}(2022){Moln{\'a}r}, {B{\'o}di}, {P{\'a}l},
  {Bhardwaj}, {Hambsch}, {Benk{\H{o}}}, {Derekas}, {Ebadi}, {Joyce},
  {Hasanzadeh}, {Kolenberg}, {Lund}, {Nemec}, {Netzel}, {Ngeow}, {Pepper},
  {Plachy}, {Prudil}, {Siverd}, {Skarka}, {Smolec}, {S{\'o}dor}, {Sylla},
  {Szab{\'o}}, {Szab{\'o}}, {Kjeldsen}, {Christensen-Dalsgaard}, \&
  {Ricker}}]{Molnáretal2022}
{Moln{\'a}r}, L., {B{\'o}di}, A., {P{\'a}l}, A., {et~al.} 2022, \apjs, 258, 8

\bibitem[{{P{\'a}l}(2012)}]{Pál2012}
{P{\'a}l}, A. 2012, \mnras, 421, 1825

\bibitem[{{P{\'a}l} {et~al.}(2020){P{\'a}l}, {Szak{\'a}ts}, {Kiss}, {B{\'o}di},
  {Bogn{\'a}r}, {Kalup}, {Kiss}, {Marton}, {Moln{\'a}r}, {Plachy},
  {S{\'a}rneczky}, {Szab{\'o}}, \& {Szab{\'o}}}]{Pál2020}
{P{\'a}l}, A., {Szak{\'a}ts}, R., {Kiss}, C., {et~al.} 2020, \apjs, 247, 26

\bibitem[{{Pedersen} {et~al.}(2021){Pedersen}, {Aerts}, {P{\'a}pics},
  {Michielsen}, {Gebruers}, {Rogers}, {Molenberghs}, {Burssens}, {Garcia}, \&
  {Bowman}}]{Pedersenetal2021}
{Pedersen}, M.~G., {Aerts}, C., {P{\'a}pics}, P.~I., {et~al.} 2021, Nature
  Astronomy, 5, 715

\bibitem[{{Plachy} {et~al.}(2021){Plachy}, {P{\'a}l}, {B{\'o}di}, {Szab{\'o}},
  {Moln{\'a}r}, {Szabados}, {Benk{\H{o}}}, {Anderson}, {Bellinger}, {Bhardwaj},
  {Ebadi}, {Gazeas}, {Hambsch}, {Hasanzadeh}, {Jurkovic}, {Kalaee}, {Kervella},
  {Kolenberg}, {Miko{\l}ajczyk}, {Nardetto}, {Nemec}, {Netzel}, {Ngeow},
  {Ozuyar}, {Pascual-Granado}, {Pilecki}, {Ripepi}, {Skarka}, {Smolec},
  {S{\'o}dor}, {Szab{\'o}}, {Christensen-Dalsgaard}, {Jenkins}, {Kjeldsen},
  {Ricker}, \& {Vanderspek}}]{Plachyetal2021}
{Plachy}, E., {P{\'a}l}, A., {B{\'o}di}, A., {et~al.} 2021, \apjs, 253, 11

\bibitem[{{Ricker} {et~al.}(2015){Ricker}, {Winn}, {Vanderspek}, {Latham},
  {Bakos}, {Bean}, {Berta-Thompson}, {Brown}, {Buchhave}, {Butler}, {Butler},
  {Chaplin}, {Charbonneau}, {Christensen-Dalsgaard}, {Clampin}, {Deming},
  {Doty}, {De Lee}, {Dressing}, {Dunham}, {Endl}, {Fressin}, {Ge}, {Henning},
  {Holman}, {Howard}, {Ida}, {Jenkins}, {Jernigan}, {Johnson}, {Kaltenegger},
  {Kawai}, {Kjeldsen}, {Laughlin}, {Levine}, {Lin}, {Lissauer}, {MacQueen},
  {Marcy}, {McCullough}, {Morton}, {Narita}, {Paegert}, {Palle}, {Pepe},
  {Pepper}, {Quirrenbach}, {Rinehart}, {Sasselov}, {Sato}, {Seager},
  {Sozzetti}, {Stassun}, {Sullivan}, {Szentgyorgyi}, {Torres}, {Udry}, \&
  {Villasenor}}]{TESS2015}
{Ricker}, G.~R., {Winn}, J.~N., {Vanderspek}, R., {et~al.} 2015, Journal of
  Astronomical Telescopes, Instruments, and Systems, 1, 014003

\bibitem[{{Riello} {et~al.}(2021){Riello}, {De Angeli}, {Evans}, {Montegriffo},
  {Carrasco}, {Busso}, {Palaversa}, {Burgess}, {Diener}, {Davidson}, {Rowell},
  {Fabricius}, {Jordi}, {Bellazzini}, {Pancino}, {Harrison}, {Cacciari}, {van
  Leeuwen}, {Hambly}, {Hodgkin}, {Osborne}, {Altavilla}, {Barstow}, {Brown},
  {Castellani}, {Cowell}, {De Luise}, {Gilmore}, {Giuffrida}, {Hidalgo},
  {Holland}, {Marinoni}, {Pagani}, {Piersimoni}, {Pulone}, {Ragaini}, {Rainer},
  {Richards}, {Sanna}, {Walton}, {Weiler}, \& {Yoldas}}]{GaiaG}
{Riello}, M., {De Angeli}, F., {Evans}, D.~W., {et~al.} 2021, \aap, 649, A3

\bibitem[{{Rimoldini} {et~al.}(2022){Rimoldini}, {Holl}, {Gavras}, {Audard},
  {De Ridder}, {Mowlavi}, {Nienartowicz}, {Jevardat de Fombelle},
  {Lecoeur-Ta{\"\i}bi}, {Karbevska}, {Evans}, {{\'A}brah{\'a}m}, {Carnerero},
  {Clementini}, {Distefano}, {Garofalo}, {Garc{\'\i}a-Lario}, {Gomel},
  {Klioner}, {Kruszy{\'n}ska}, {Lanzafame}, {Lebzelter}, {Marton}, {Mazeh},
  {Molinaro}, {Panahi}, {Raiteri}, {Ripepi}, {Szabados}, {Teyssier},
  {Trabucchi}, {Wyrzykowski}, {Zucker}, \& {Eyer}}]{rimoldini-2022}
{Rimoldini}, L., {Holl}, B., {Gavras}, P., {et~al.} 2022, arXiv e-prints,
  arXiv:2211.17238

\bibitem[{{Ripepi} {et~al.}(2022){Ripepi}, {Clementini}, {Molinaro}, {Leccia},
  {Plachy}, {Moln{\'a}r}, {Rimoldini}, {Musella}, {Marconi}, {Garofalo},
  {Audard}, {Holl}, {Evans}, {Jevardat de Fombelle}, {Lecoeur-Taibi},
  {Marchal}, {Mowlavi}, {Muraveva}, {Nienartowicz}, {Sartoretti}, {Szabados},
  \& {Eyer}}]{Ripepietal2022}
{Ripepi}, V., {Clementini}, G., {Molinaro}, R., {et~al.} 2022, arXiv e-prints,
  arXiv:2206.06212

\bibitem[{{Ripepi} {et~al.}(2019){Ripepi}, {Molinaro}, {Musella}, {Marconi},
  {Leccia}, \& {Eyer}}]{Ripepietal2019}
{Ripepi}, V., {Molinaro}, R., {Musella}, I., {et~al.} 2019, \aap, 625, A14

\bibitem[{{Simon} \& {Lee}(1981)}]{simon-lee-1981}
{Simon}, N.~R. \& {Lee}, A.~S. 1981, \apj, 248, 291

\bibitem[{{Skarka} {et~al.}(2022){Skarka}, {{\v{Z}}{\'a}k}, {Fedurco},
  {Paunzen}, {Henzl}, {Ma{\v{s}}ek}, {Karjalainen}, {Sanchez Arias},
  {S{\'o}dor}, {Auer}, {Kab{\'a}th}, {Karjalainen}, {Li{\v{s}}ka}, \&
  {{\v{S}}tegner}}]{Skarkaetal2022}
{Skarka}, M., {{\v{Z}}{\'a}k}, J., {Fedurco}, M., {et~al.} 2022, \aap, 666,
  A142

\bibitem[{{Soszy{\'n}ski} {et~al.}(2008){Soszy{\'n}ski}, {Poleski}, {Udalski},
  {Szyma{\'n}ki}, {Kubiak}, {Pietrzy{\'n}ski}, {Wyrzykowski}, {Szewczyk}, \&
  {Ulaczyk}}]{Soszynskietal2008}
{Soszy{\'n}ski}, I., {Poleski}, R., {Udalski}, A., {et~al.} 2008, \actaa, 58,
  163

\bibitem[{{Sterken} \& {Jerzykiewicz}(1993)}]{SterkenandJerzykiewicz1993}
{Sterken}, C. \& {Jerzykiewicz}, M. 1993, \ssr, 62, 95

\bibitem[{{Szab{\'o}} {et~al.}(2009){Szab{\'o}}, {Koll{\'a}th}, {Moln{\'a}r},
  {Benk{\H{o}}}, \& {Szabados}}]{Szaboetal2009}
{Szab{\'o}}, R., {Koll{\'a}th}, Z., {Moln{\'a}r}, L., {Benk{\H{o}}}, J.~M., \&
  {Szabados}, L. 2009, in American Institute of Physics Conference Series, Vol.
  1170, Stellar Pulsation: Challenges for Theory and Observation, ed. J.~A.
  {Guzik} \& P.~A. {Bradley}, 102--104

\bibitem[{{Szab{\'o}} {et~al.}(2021){Szab{\'o}}, {K{\'o}sp{\'a}l},
  {{\'A}brah{\'a}m}, {Park}, {Siwak}, {Green}, {Mo{\'o}r}, {P{\'a}l},
  {Acosta-Pulido}, {Lee}, {Cseh}, {Cs{\"o}rnyei}, {Hanyecz},
  {K{\"o}nyves-T{\'o}th}, {Krezinger}, {Kriskovics}, {Ordasi}, {S{\'a}rneczky},
  {Seli}, {Szak{\'a}ts}, {Szing}, \& {Vida}}]{SzabóZs-2021}
{Szab{\'o}}, Z.~M., {K{\'o}sp{\'a}l}, {\'A}., {{\'A}brah{\'a}m}, P., {et~al.}
  2021, \apj, 917, 80

\bibitem[{{Szab{\'o}} {et~al.}(2022){Szab{\'o}}, {K{\'o}sp{\'a}l},
  {{\'A}brah{\'a}m}, {Park}, {Siwak}, {Green}, {P{\'a}l}, {Acosta-Pulido},
  {Lee}, {Ibrahimov}, {Grankin}, {Kov{\'a}cs}, {Bora}, {B{\'o}di}, {Cseh},
  {Cs{\"o}rnyei}, {Dr{\'o}{\.z}d{\.z}}, {Hanyecz}, {Ign{\'a}cz}, {Kalup},
  {K{\"o}nyves-T{\'o}th}, {Krezinger}, {Kriskovics}, {Og{\l}oza}, {Ordasi},
  {S{\'a}rneczky}, {Seli}, {Szak{\'a}ts}, {S{\'o}dor}, {Szing}, {Vida}, \&
  {Vink{\'o}}}]{SzabóZs-2022}
{Szab{\'o}}, Z.~M., {K{\'o}sp{\'a}l}, {\'A}., {{\'A}brah{\'a}m}, P., {et~al.}
  2022, \apj, 936, 64

\bibitem[{{Waelkens}(1991)}]{Waelkens1991}
{Waelkens}, C. 1991, \aap, 246, 453

\bibitem[{{Waelkens}(1996)}]{Waelkens1996}
{Waelkens}, C. 1996, \aap, 311, 873

\bibitem[{{Wood}(1976)}]{Woodetal1976}
{Wood}, P.~R. 1976, \mnras, 174, 531

\bibitem[{{Zsidi} {et~al.}(2022{\natexlab{a}}){Zsidi}, {Fiorellino},
  {K{\'o}sp{\'a}l}, {{\'A}brah{\'a}m}, {B{\'o}di}, {Hussain}, {Manara}, \&
  {P{\'a}l}}]{zsidi2-2022}
{Zsidi}, G., {Fiorellino}, E., {K{\'o}sp{\'a}l}, {\'A}., {et~al.}
  2022{\natexlab{a}}, \apj, 941, 177

\bibitem[{{Zsidi} {et~al.}(2022{\natexlab{b}}){Zsidi}, {Manara},
  {K{\'o}sp{\'a}l}, {Hussain}, {{\'A}brah{\'a}m}, {Alecian}, {B{\'o}di},
  {P{\'a}l}, \& {Sarkis}}]{zsidi1-2022}
{Zsidi}, G., {Manara}, C.~F., {K{\'o}sp{\'a}l}, {\'A}., {et~al.}
  2022{\natexlab{b}}, \aap, 660, A108

\end{thebibliography}

\begin{appendix}
\section{CoRoT and TESS photometry}
\label{app:photometry}

Tables\,\ref{table:app_corot} and \ref{table:app_TESS} contain CoRoT and TESS (sector 06 and 33) photometric data, as described in Sections~\ref{sect:corot} and \ref{sect:tess}. Both tables are available online in machine-readable format.

\begin{table*}
\caption{Sample table of CoRoT IRa01 run measurements in $W$ from the Faint Star Catalogue.}            
\label{table:app_corot}      
\centering            
\begin{tabular}{c c c c c c c}
\hline\hline            

\smallskip CoRoT ID & BJD   & $W$  & $\sigma_W$  & run\\
& (days) & (mag) & (mag) &  \\
\hline 
\noalign{\smallskip}
102835671   & 2454135.05334 &   15.3388 &       0.0061 &        IRa01 \\
102835671   & 2454135.05926     &   15.3327 &   0.0106 &        IRa01 \\
102835671       & 2454135.06519 &       15.3345 &       0.0068 &        IRa01 \\
102835671       & 2454135.07111 &       15.3375 &       0.0085 &        IRa01 \\
102835671       & 2454135.08296 &       15.3332 &       0.0077 &        IRa01 \\
102835671       & 2454135.08889 &       15.3290 &       0.0118 &        IRa01 \\
102835671       & 2454135.09482 &       15.3331 &       0.0056 &        IRa01 \\
102835671       & 2454135.10074 &       15.3363 &       0.0057 &        IRa01 \\
\multicolumn{5}{l}{$\dots$}\\
\hline                                  
\end{tabular}

\tablefoot{
 The table contains CoRoT IDs, the barycentric Julian date of the measurements, white magnitudes, the error of white magnitudes, and the name of the run. The entire table is available  online in machine-readable format.}
\end{table*}

\begin{table*}
\caption{Sample table of the TESS magnitudes in sectors 06 and 33, based on the \texttt{FITSH} differential image photometry of the FFIs of each ULA Cepheid candidate.}            
\label{table:app_TESS}      
\centering            
\begin{tabular}{c c c c c c c}
\hline\hline            

\smallskip CoRoT ID & BJD   & TESS mag  & $\sigma_\mathrm{TESS\,mag}$  & sector\\
& (days) & (mag) & (mag) &  \\
\hline 
\noalign{\smallskip}
102835671   &   2458468.30264  &        14.9005   &     0.0265    &     TESS06 \\
102835671   &   2458468.32348  &        14.8886   &     0.0233    &     TESS06 \\
102835671   &   2458468.34431  &        14.8758   &     0.0251    &     TESS06 \\
102835671   &   2458468.36514  &        14.8840   &     0.0231    &     TESS06 \\
102835671   &   2458468.38598  &        15.0000   &     0.0524    &     TESS06 \\
102835671   &   2458468.40681  &        14.8770   &     0.0228    &     TESS06 \\
102835671   &   2458468.42764  &        14.8834   &     0.0209    &     TESS06 \\
102835671   &   2458468.44848  &        14.9034   &     0.0220    &     TESS06 \\
\multicolumn{5}{l}{$\dots$}\\
\hline                                  
\end{tabular}

\tablefoot{
 The table contains CoRoT IDs, the barycentric Julian date of the measurements, TESS magnitudes, the error of TESS magnitudes, and the number of the TESS sector. The entire table is available  online in machine-readable format.}
\end{table*}

\section{Gaia DR3 parallax versus geometric distances}

\begin{figure*}
    \centering
    \includegraphics[width=\columnwidth]{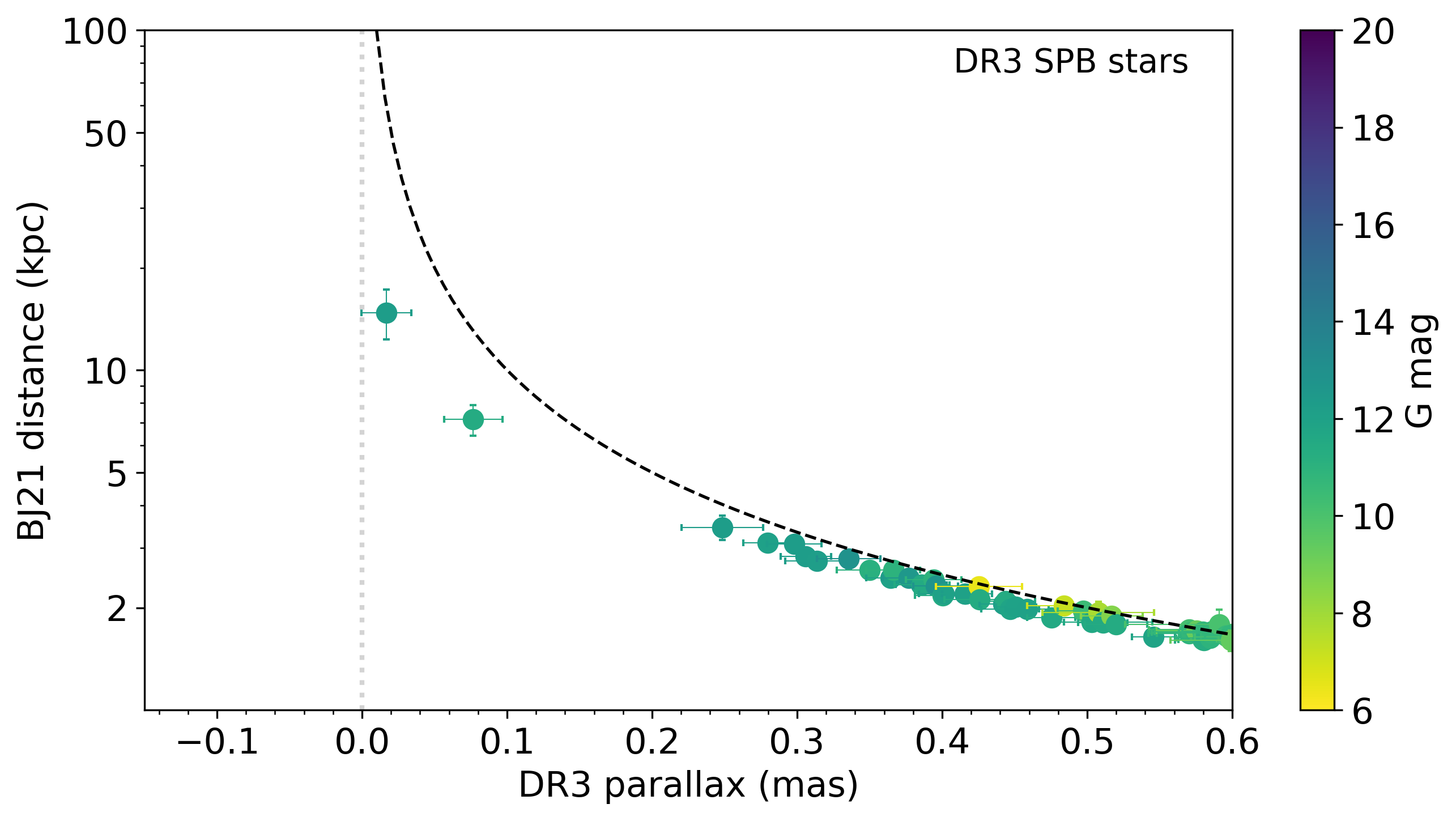}
    \includegraphics[width=\columnwidth]{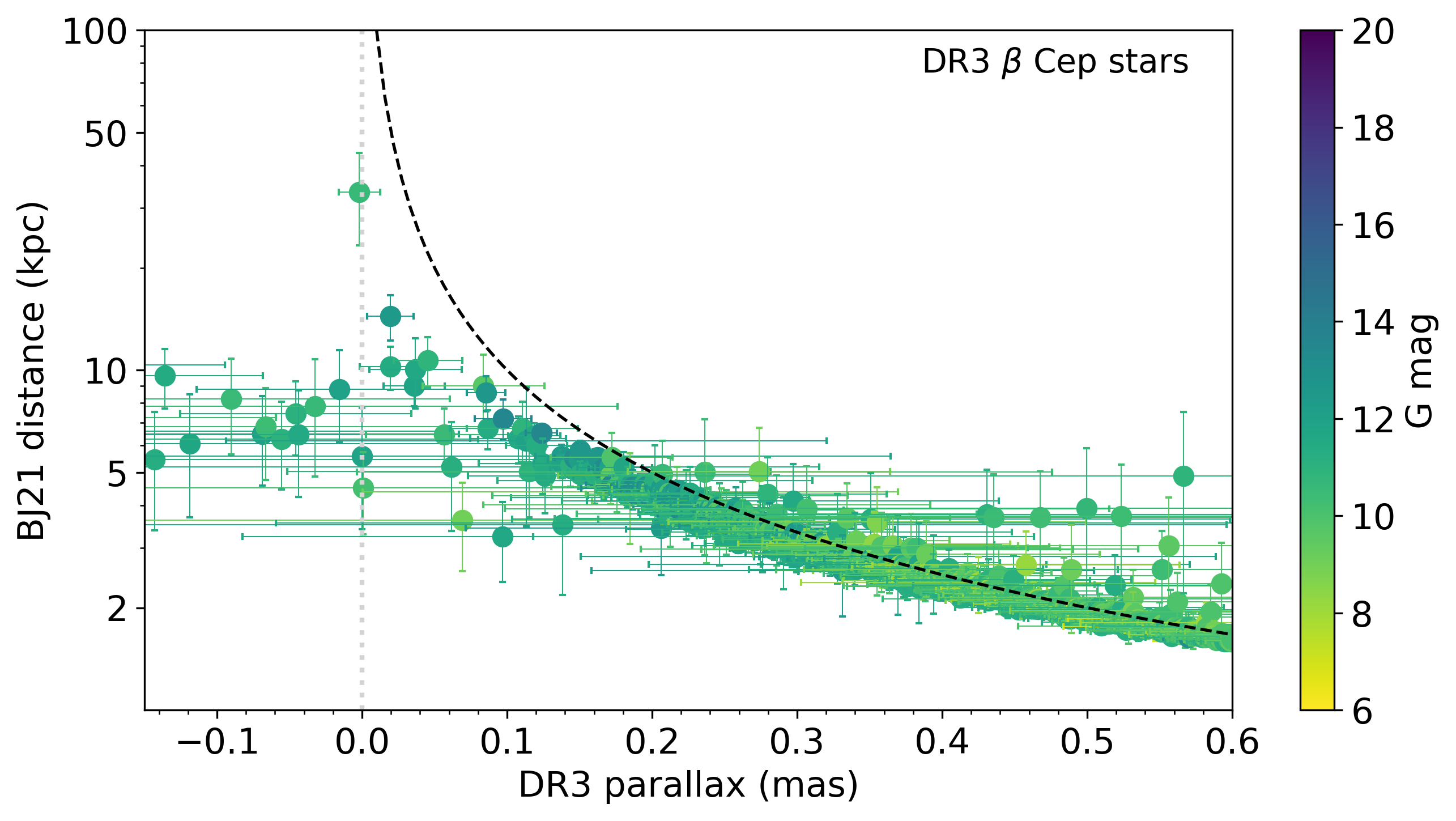}
    \includegraphics[width=\columnwidth]{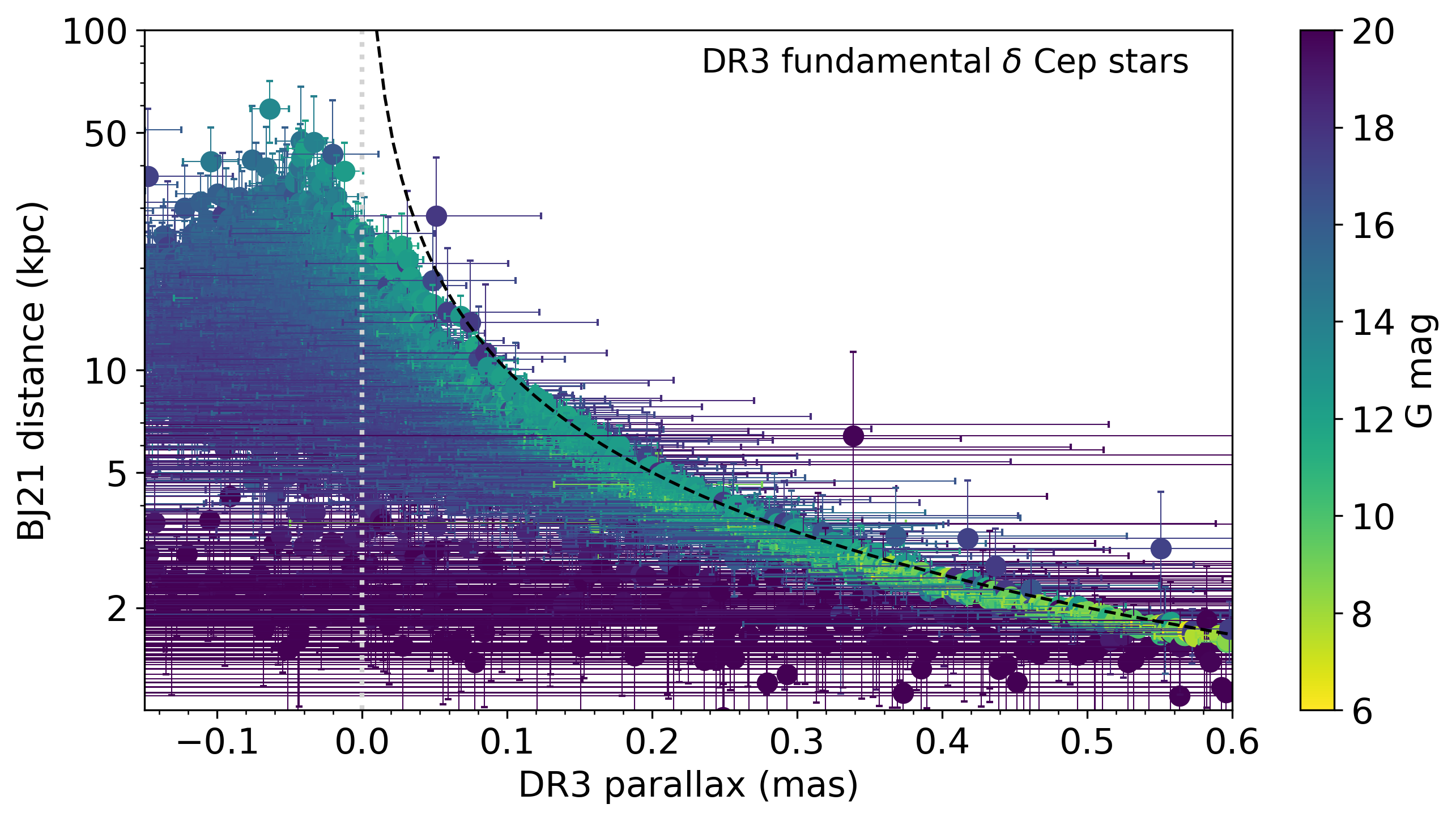}
    \includegraphics[width=\columnwidth]{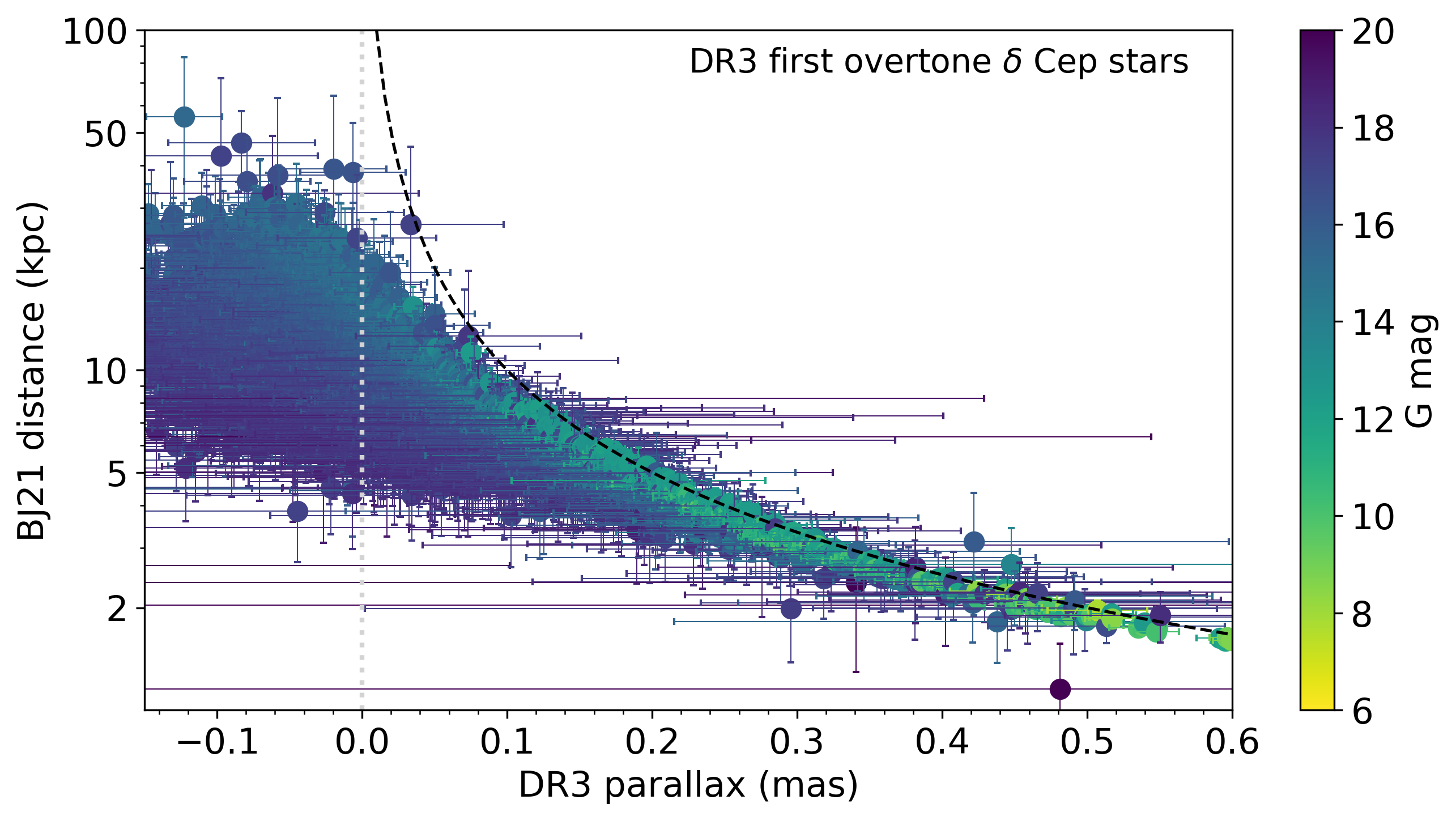}
    \caption{Gaia DR3 parallaxes ($\varpi$) versus geometric distances calculated by \citet{BailerJones2021} for slowly pulsating B (SPB), $\beta$ Cephei, as well as fundamental and first overtone $\delta$ Cephei stars with the measured and calculated errorbars for the whole sample. The G-band brightness of the stars are color coded. One can see that the calculated distance for the closest stars is overestimated, as it does not follow the $\rho = 1/\varpi$ parallax-distance inversion law (black dashed line). Although, the turning point from the parallax-distance inversion law differs for each star type.}
    \label{fig:plxover}
\end{figure*}

Figure\,\ref{fig:plxover} shows Gaia DR3 parallaxes versus the calculated geometric distances of \citet{BailerJones2021} of SPB stars (upper left), $\beta$ Cephei (upper right), as well as fundamental and first-overtone Cepheids (lower panels).
The black dashed line presents the simple $\rho=1/\varpi$ parallax--distance law.
It can be seen that the brightest stars follow this law very closely, but not perfectly. 
That is because Gaia produces a probability distribution for the parallax from the measurements instead of a single numerical value, and then that distribution gets inverted. 
The posterior probability distribution of the distance, and the highest likelihood value thus depend on the shape of the input distribution plus the chosen prior. 
The prior approximates the density of stars within the Milky Way and prevents them from appearing at unphysically large distances. 
This also means that stars with uninformative parallax data preferentially appear near the median distance of the prior.
This can be observed in the distribution of stars in Fig.~\ref{fig:plxover}.

Stars that have large uncertainties ($\sigma_\varpi$) and/or negative parallaxes are the faintest (e.g., $ G \gtrsim 15-16$\,mag in the case of $\delta$ Cepheids).
To avoid large parallax uncertainties, we filtered the data with the parallax-over-error ratio ($\varpi/\sigma_\varpi$, listed as \texttt{parallax\_over\_error} in the Gaia archive).
Stars that have $\varpi/\sigma_\varpi>5$ (a threshold of $20\%$ statistical error in parallax data) are not expected to be considered in further investigations. 

\begin{figure*}
    \centering
    \includegraphics[width=\columnwidth]{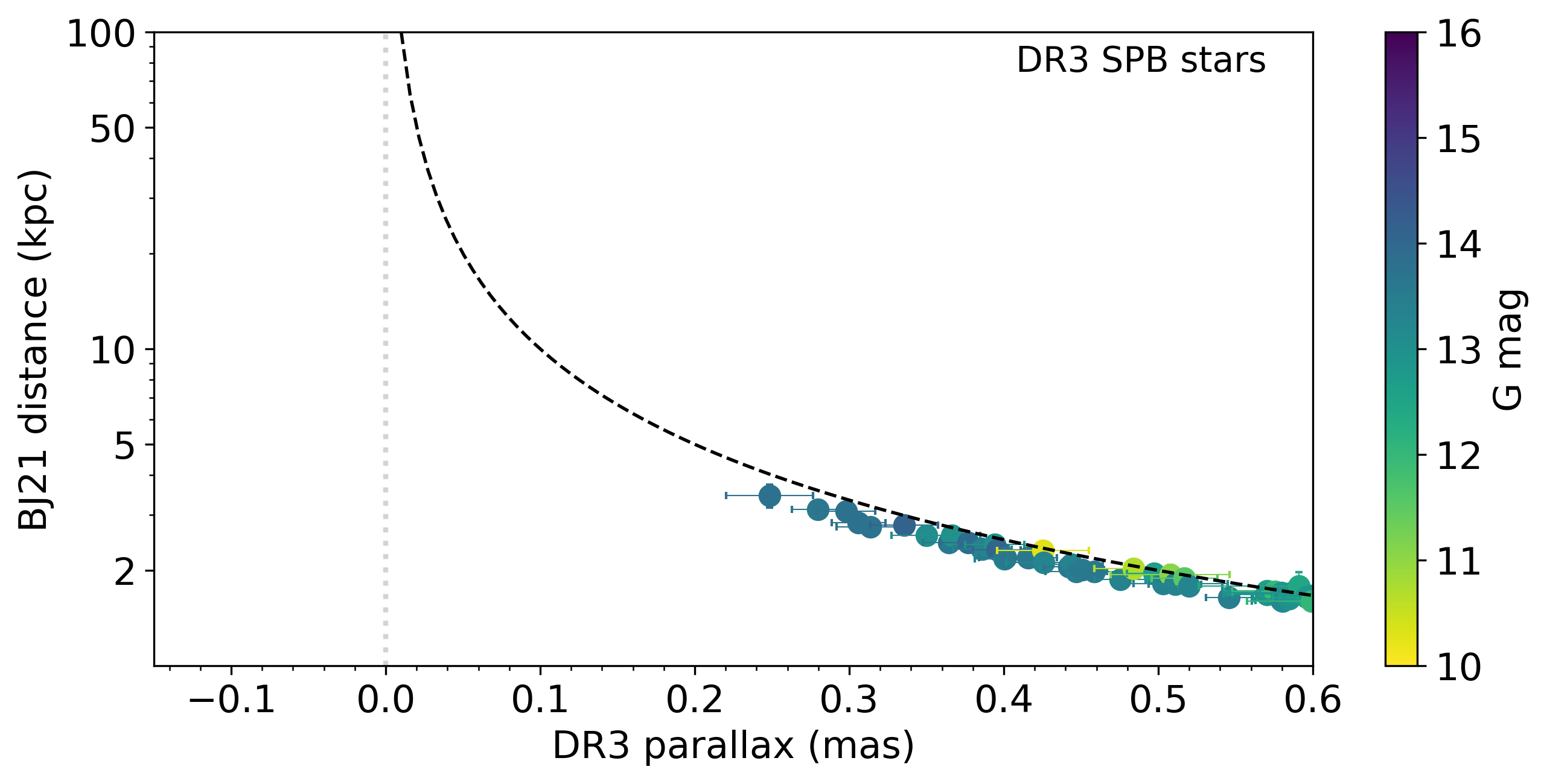}
    \includegraphics[width=\columnwidth]{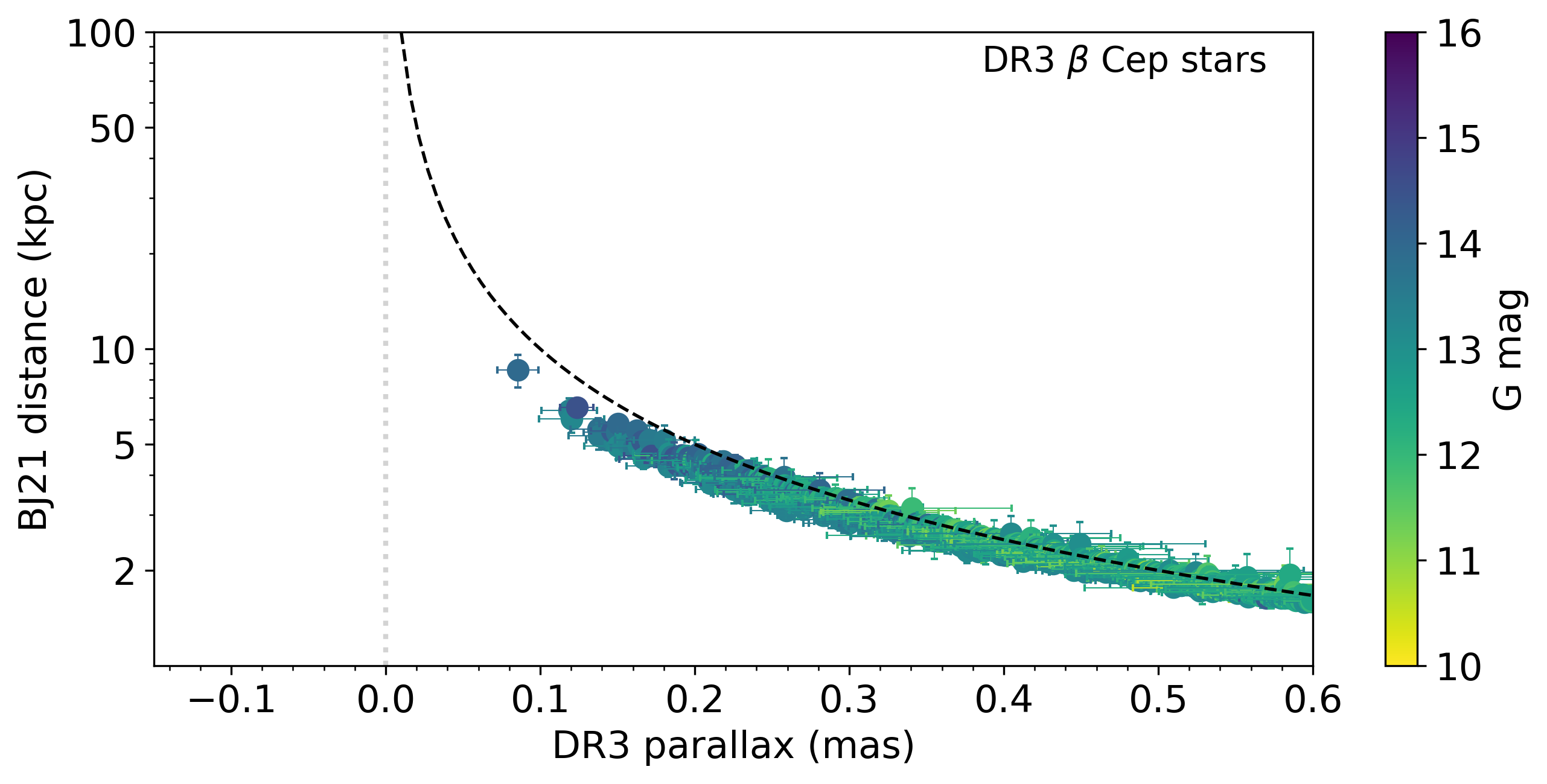}
    \includegraphics[width=\columnwidth]{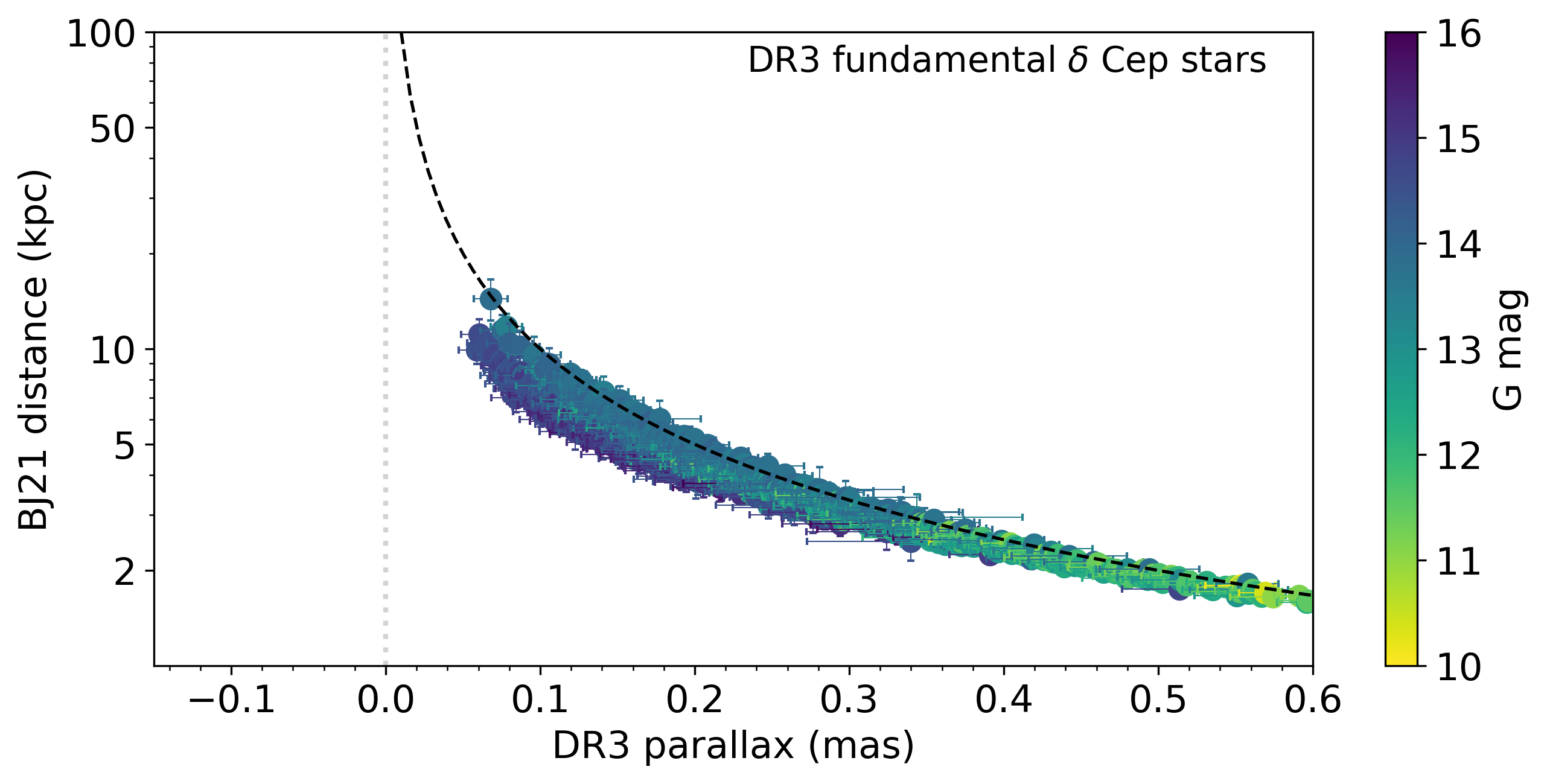}
    \includegraphics[width=\columnwidth]{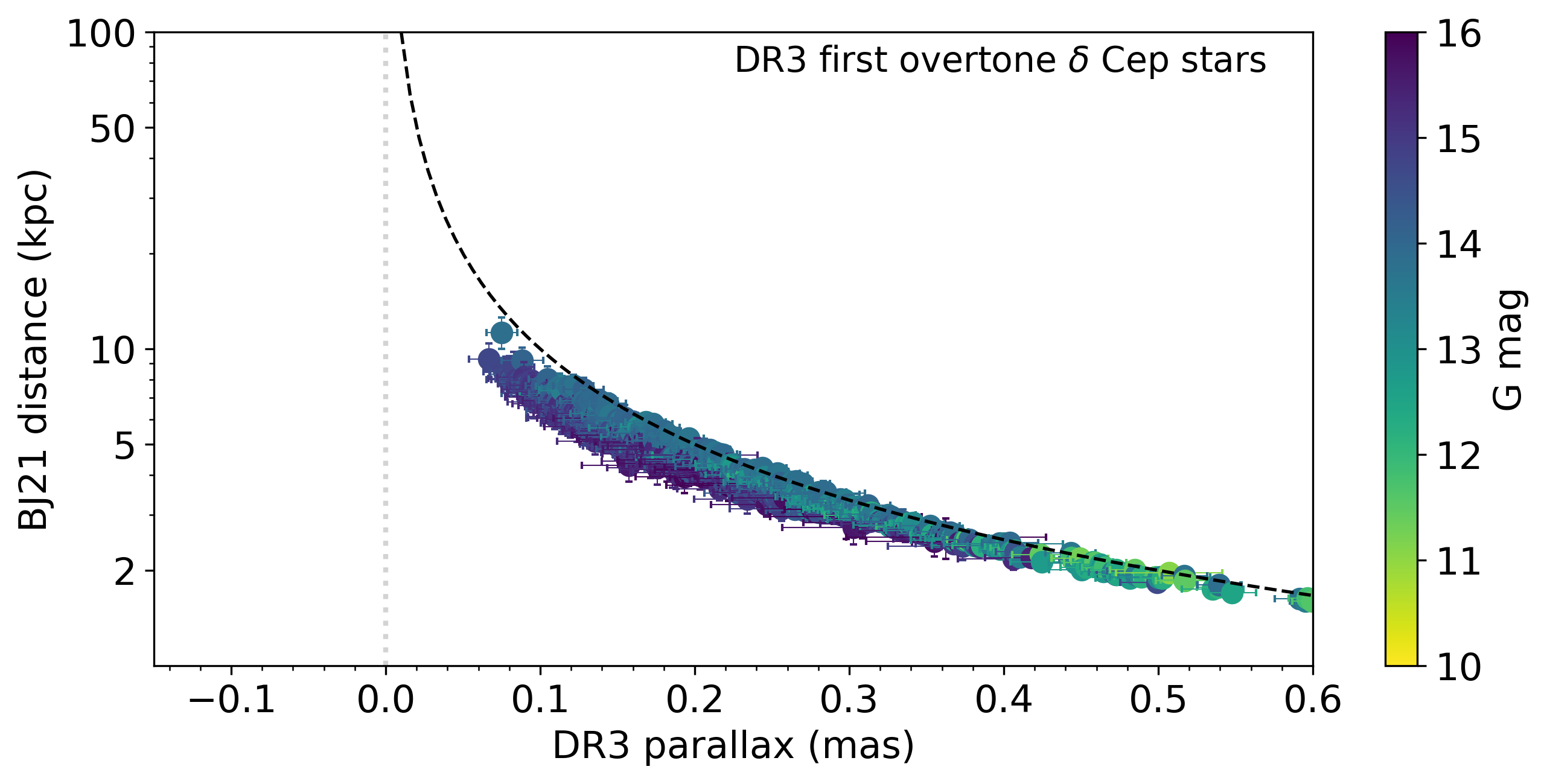}
    \caption{Same as Figure~\ref{fig:plxover}, but for the filtered sample with the $\varpi/\sigma_\varpi > 5$ threshold (see details in Section~\ref{sec:distances}).}
    \label{fig:plxover_cut}
\end{figure*}

Figure\,\ref{fig:plxover_cut} is the same as Figure\,\ref{fig:plxover} but for the filtered sample, where only stars that meet the threshold of $\varpi / \sigma_\varpi > 5$ are shown. 
These selections result in samples that follow the parallax-distance law much more closely. 

\section{Infrared CMDs}

Figure\,\ref{camdIR} shows infrared CMDs, based on 2MASS $J$, $H$, and $K$ data. Brightnesses are converted to dereddened absolute magnitudes in the same way, as described in \ref{subsec:gaiadata}. 
They show the same structure as the Gaia CMD in Section~\ref{sect:gaiacmd}, except that the reddening vectors are shorter and flatter in these passbands. 
Two stars, 4858 and 7671, are main-sequence dwarfs that suffer only low absorption. 
The other four stars move to the blue giant part of the main sequence after corrections.

\begin{figure*}
    \centering
    \includegraphics[width=\textwidth]{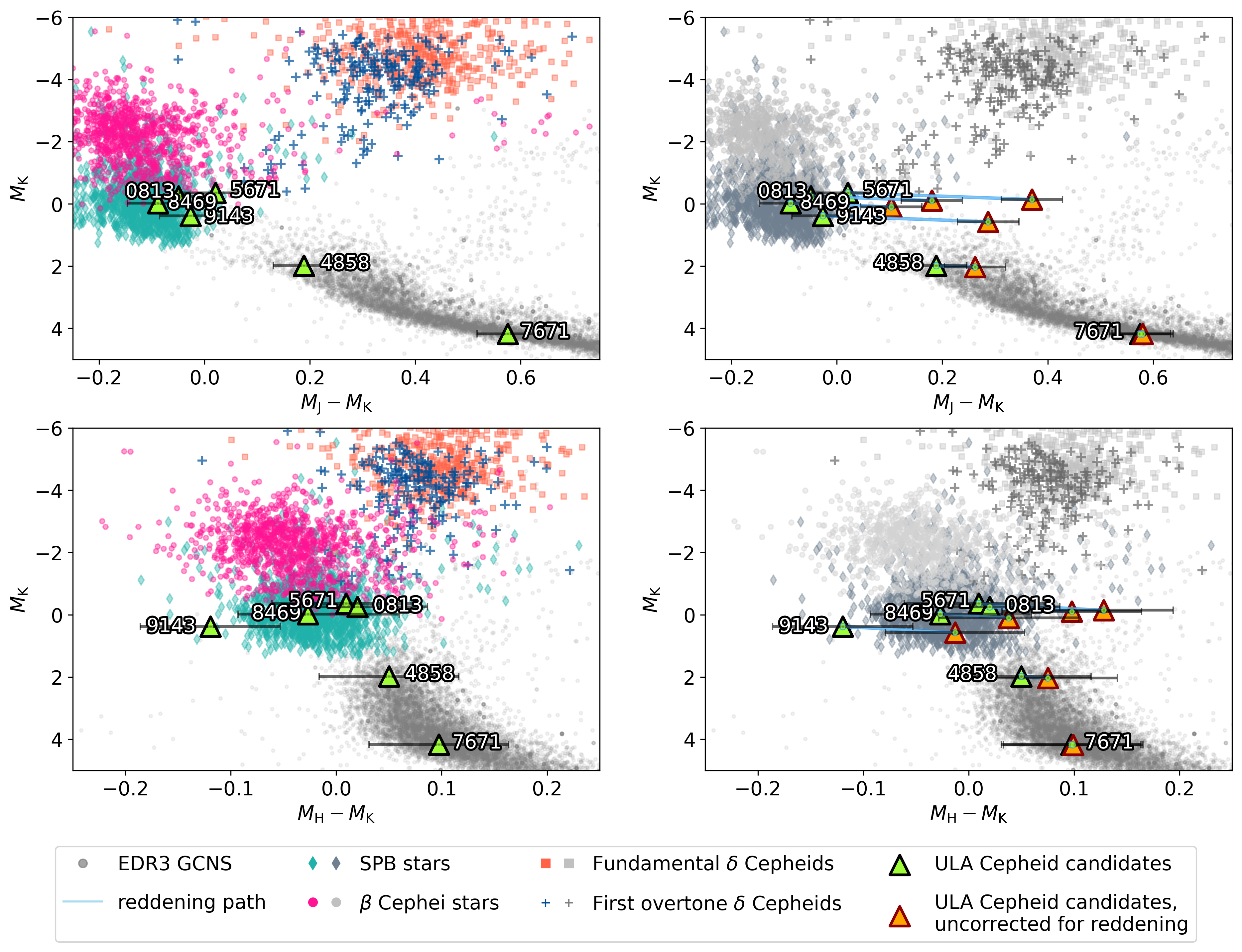}
    \caption{CMDs in the infrared, based on 2MASS data. 
    The stars are shown in the same way as in Figure\,\ref{fig:camd}.}
    \label{fig:camdIR}
\end{figure*}

\end{appendix}

\end{document}